\documentclass[a4paper,12pt]{article}
\usepackage{epsfig}
\usepackage{amssymb}
\usepackage{cite}
\usepackage{t1enc}
\usepackage{epstopdf}
\usepackage{graphicx}
\usepackage[latin1]{inputenc}
\usepackage[english]{babel}
\usepackage{slashed}
\usepackage{axodraw}
\usepackage{mathabx}
\usepackage{multicol}
\usepackage{slashed}
\usepackage{mathabx}
\usepackage{multicol}

\declareslashed{}{/}{-0.004}{0}{R}
\newcommand{\Lag}{\mathcal{L}}

\newcommand{\cw}{c_w}
\newcommand{\sw}{s_w}

\newcommand{\isqt}{\frac{1}{\sqrt{2}}}
\newcommand{\YU}[1]{\left( Y_U   \right)^{\phantom{*}}_{#1}}

\newcommand{\lam}[1]{ \lambda^{\phantom{*}}_{#1} }

\newcommand{\lamp}[1]{ {\lambda'}^{\phantom{*}}_{\! \!  #1} }

\newcommand{\uL}[3]{ u^{#2 #3}_{L #1} }
\newcommand{\dL}[3]{ d^{#2 #3}_{L #1} }
\newcommand{\uR}[3]{ u^{#2 #3}_{R #1} }

\newcommand{\dR}[3]{ d^{#2 #3}_{R #1} }

\newcommand{\seL}[1]{ \tilde{e}^{\phantom{*}}_{L #1}}

\newcommand{\snuL}[1]{ \tilde{\nu}^{\phantom{*}}_{L #1} }

\newcommand{\shbz}{ h_2^0 }

\newcommand{\shbp}{ h_2^+ }

\newcommand{\ZH}[1]{ Z_{H#1} }
\newcommand{\ZR}[1]{ Z_{R#1} }
\newcommand{\ZA}[1]{ Z_{A#1} }
\newcommand{\Zp}[1]{ Z_{+#1} }

\newcommand{\ZN}[1]{ Z_{N#1} }
\newcommand{\Zsu}[1]{ Z_{\tilde{u}#1} }
\newcommand{\Zsd}[1]{ Z_{\tilde{d}#1} }

\newcommand{\ZdR}[1]{ Z_{d_R#1} }
\newcommand{\ZuL}[1]{ Z_{u_L#1} }

\newcommand{\ZHs}[1]{ Z^*_{H#1} }

\newcommand{\Zms}[1]{ Z^*_{-#1} }

\newcommand{\Zsus}[1]{ Z^*_{\tilde{u}#1} }
\newcommand{\Zsds}[1]{ Z^*_{\tilde{d}#1} }
\newcommand{\ZdLs}[1]{ Z^*_{d_L#1} }

\newcommand{\ZuRs}[1]{ Z^*_{u_R#1} }
\newcommand{\Hp}[1]{ H^+_{ #1 }}
\newcommand{\Hz}[1]{ H^0_{ #1 } }
\newcommand{\Az}[1]{ A^0_{ #1 } }
\newcommand{\Kp}[2]{ \kappa^{+#2}_{#1} }
\newcommand{\Km}[2]{ \kappa^{-#2}_{#1} }
\newcommand{\Kz}[2]{ \kappa^{0#2}_{#1} }
\newcommand{\su}[2]{ \tilde{u}_{#1}^{#2} }
\newcommand{\sd}[2]{ \tilde{d}_{#1}^{#2} }

\newcommand{\Hps}[1]{ H^{+*}_{ #1 }}

\newcommand{\sus}[2]{ \tilde{u}_{#1}^{*#2} }
\newcommand{\sds}[2]{ \tilde{d}_{#1}^{*#2} }

\usepackage{axodraw}

\newcommand{\iin}[8]
{ \begin{tabular}{ll}
  \begin{picture}(150,70)(0,-10)
    \DashLine(60,0)(10,0){5} \Text(0,0)[c]{$#3$} \ArrowLine(110,0)(60,0) \Text(120,0)[c]{$#2$}
    \ArrowLine(60,50)(60,0) \Text(65,45)[l]{$#1$} \Vertex(60,0){2}
  \end{picture} & \raisebox{5\unitlength}
  { \begin{minipage}{5cm} \lefteqn{ #4 } \lefteqn{ #5 } \lefteqn{ #6 } \lefteqn{ #7 } \lefteqn{ #8 } \end{minipage} }
\end{tabular} }

\newcommand{\iisi}[8]
{ \begin{tabular}{ll}
  \begin{picture}(150,70)(0,-10)
    \DashArrowLine(10,0)(60,0){5} \Text(0,0)[c]{$#3$} \ArrowLine(110,0)(60,0) \Text(120,0)[c]{$#2$}
    \ArrowLine(60,50)(60,0) \Text(65,45)[l]{$#1$} \Vertex(60,0){2}
  \end{picture} & \raisebox{15\unitlength}
  { \begin{minipage}{5cm} \lefteqn{ #4 } \lefteqn{ #5 } \lefteqn{ #6 } \lefteqn{ #7 } \lefteqn{ #8 }\end{minipage} }
\end{tabular} }

\newcommand{\iiso}[8]
{ \begin{tabular}{ll}
  \begin{picture}(150,70)(0,-10)
    \DashArrowLine(60,0)(10,0){5} \Text(0,0)[c]{$#3$} \ArrowLine(110,0)(60,0) \Text(120,0)[c]{$#2$}
    \ArrowLine(60,50)(60,0) \Text(65,45)[l]{$#1$} \Vertex(60,0){2}
  \end{picture} & \raisebox{15\unitlength}
  { \begin{minipage}{5cm} \lefteqn{ #4 } \lefteqn{ #5 } \lefteqn{ #6 } \lefteqn{ #7 } \lefteqn{ #8 }\end{minipage} }
\end{tabular} }

\newcommand{\be}{\begin{equation}}
\newcommand{\ee}{\end{equation}}
\newcommand{\br}{\begin{eqnarray}}
\newcommand{\er}{\end{eqnarray}}

\newcommand{\LMSSM}{{\slash\hspace*{-0.24cm}}L-MSSM}

\begin{document}
\tolerance=100000

\begin{flushright}
IPPP/06/71\\[-1mm]
DCPT/06/142\\[-1mm]
\end{flushright}

\bigskip

\begin{center}
{\Large \bf $l \rightarrow l' \gamma$ in the Lepton Number Violating MSSM}\\[1.7cm] 
\large Steven Rimmer \\[3mm]
\normalsize {\it Institute for Particle Physics Phenomenology (IPPP), Durham 
DH1 3LE, UK } \\[2mm]
\end{center}

\vspace*{0.8cm}\centerline{\bf ABSTRACT}
\vspace{0.1cm}\noindent{\small
A minimal particle content supersymmetric model with a discrete ${\cal Z}_3$ symmetry,
allowing lepton number violating terms, is studied.  Within this model, the
neutrino masses and mixing can be generated by lepton number
violating couplings.  Choosing parameters which correctly
describe both the masses and mixing in the neutrino sector, we consider 
their repercussions in flavour violating radiative
lepton decays, $l \rightarrow l' \gamma$.  Such decays have not been observed 
and, accordingly, soon to be improved, upper bounds on their
branching ratios exist.  We do not assume
dominance of either the bilinear or trilinear couplings.  We note that
certain parameter sets, which correctly describe the neutrino sector,
will also generate observable branching ratios, some of which are already precluded,
and suggest four such sets as Benchmark scenarios. 
}
\vspace*{\fill}
\newpage
\setcounter{equation}{0}
\setcounter{figure}{0}
\renewcommand{\thefigure}{\arabic{figure}}
\section{Introduction}
\label{sec:intro}
Processes which do not conserve lepton flavour, the flavour oscillations
in the neutrino sector, have been observed~\cite{Schwetz:2006dh}.  This is in contrast with the charged
sector, where no such observation has been made. 
The decays $\tau \rightarrow \mu \gamma, e \gamma$ and $\mu \rightarrow e \gamma$ 
can be driven solely by the known
lepton flavour violation in the neutral sector, the branching ratio
will be small~\cite{Cheng:1976uq}, however, well below current experimental limits, due to the magnitude of
the neutrino mass.  Noticing that in many extensions of the Standard Model this branching
ratio increases greatly,\footnote{
The effect of lepton flavour non-conservation from the charged slepton mass matrix in supersymmetric 
extensions of the Standard Model where a seesaw mechanism results in light Majorana neutrinos is 
noted in Ref.~\cite{Borzumati:1986qx}, this work is extended in Ref~\cite{deCarlos:1995ah}, where 
bounds for off-diagonal terms are calculated.  In Ref.~\cite{Hisano:2001qz} the results for this model are 
correlated with neutrino masses and the $(g_\mu-2)$ data and in Ref.~\cite{Lavignac:2001vp} the possibility
of discriminating between different supersymmetric seesaw models is investigated.  A bottom-up approach
is considered in Ref.\cite{Casas:2001sr}, resulting in predictions for the $\mu \rightarrow e \gamma$ branching ratio.
Methods for discerning models with heavy right handed neutrinos from R-parity violating models using a number of decays
are studied in Ref.~\cite{deGouvea:2000cf}.  Renormalisation group effects due to R-parity violating couplings
 and their effect on the $\mu \rightarrow e \gamma$ branching ratio are considered
in Ref.~\cite{deCarlos:1996du}.
} together with the fact that experimental bounds for these decays 
will soon be improved by several orders of magnitude, suggests that these decays are a 
valuable place 
to scrutinise the Standard Model and test theories which extend it.   

The model considered here, the Lepton
Number Violating Minimal Supersymmetric Standard Model (\LMSSM), 
is a well motivated extension of the Standard
Model.  When constructing a supersymmetric
model dangerous operators arise which, unless highly suppressed, would give
rise to an unacceptable rate for proton decay.  As such, a discrete symmetry
must be imposed when constructing the Lagrangian~\cite{Ibanez:1991pr,Dreiner:2005rd}.
We choose a discrete symmetry which allows lepton number violating terms, but
not baryon number violating terms, resulting in proton stability.  The lepton
number violating terms violate R-parity.  The superpotential for the \LMSSM~is given by
\begin{eqnarray}
{\cal W}_{\rm ~ \LMSSM} = \epsilon_{ab}\left( \frac{1}{2}
\lambda_{\alpha\beta k}{\cal L}_{\alpha}^a {\cal L}_{\beta}^b
\bar{E}_k \: + \: \lambda'_{\alpha jk} {\cal L}_{\alpha}^a Q_j^{b,x}
\bar{D}_{k,x} \: + \: (Y_U)_{jk} Q_j^{a,x} H_2^b \bar{U}_{k,x} \: - \:
\mu_\alpha {\cal L}_\alpha^a H_2^b \right ),\nonumber \\
\label{superpot1}
\end{eqnarray}
where $Q_i^{a\,x},\;{\bar D}_i^x,\;{\bar U}_i^x,\; {\cal
L}_i^a,\;{\bar E}_i,\; H_1^a,\;H_2^a$ are the chiral superfield
particle content, $i=1,2,3$ is a generation index, $x=1,2,3$ and
$a=1,2$ are $SU(3) $ and $SU(2)$ gauge indices, respectively.  
${\cal L}^a_{\alpha=0,\ldots,3}=(H_1^a,\,L_{i=1,2, 3}^a)$ and hence,
$\lambda_{0jk}$ are the lepton Yukawa couplings and $\lambda_{ijk}$ are lepton
number violating parameters, whose role will be considered later.
$\mu_\alpha$ is the generalised dimensionful $\mu$-parameter, with
$\mu_0$ and $\mu_i$, $i=1,...3$ the lepton number conserving and
violating parts respectively.  Similarly, $\lambda'_{0 j k}$ are the Yukawa
couplings for the down-type quarks and $\lambda'_{ijk}$ violate lepton
number.  Finally, $(Y_U)_{ij}$ are Yukawa matrices with
$\epsilon_{ab}$ being the totally anti-symmetric tensor
$\epsilon_{12}=+1$.  The are further sources of lepton number violation in the
soft breaking sector, however, we present here merely the terms which will play
a role in this paper,  
\begin{equation} 
\Lag_{\rm SSB} \supset
B_\alpha \snuL{\alpha} \shbz 
- B_\alpha \seL{\alpha} \shbp \; \;+\; \textup{H. c.} \; \;, 
\end{equation}
where $B_\alpha$ is the four-component bilinear term $B_\alpha = (B_0,
B_i)$ are $B_i$ are the lepton number violating components.
The full supersymmetry breaking part of the Lagrangian
and the mass matrices along with further details of the model are 
presented in Refs.~\cite{Allanach:2003eb,Dedes:2006ni}.  The calculation is
performed in a basis where the vacuum expectation values for the 
sneutrinos are zero; a method for moving to this basis from the most general scalar
potential and properties of this basis are outlined in Ref.~\cite{Dedes:2005ec}.

A particularly noteworthy feature is that the current experimental values 
of neutrino mass squared differences and mixing can be accommodated within the model, being 
determined by the value of lepton number violating couplings in either the superpotential,
or the supersymmetry breaking terms of the Lagrangian~\cite{Dedes:2006ni}.
In fact it has been shown that, one, and only one, neutrino mass can be generated at 
tree-level~\cite{Hall:1983id,Joshipura:1994ib,Banks:1995by,Nowakowski:1995dx} with 
the masses of the two remaining generations arising through radiative corrections, 
producing the hierarchy between solar and atmospheric mass differences in a convenient fashion.  
Furthermore, this scenario can arise from a Froggatt-Nielsen model in which the
discrete symmetry is due to the breaking of a $U(1)_X$~\cite{Dreiner:2006xw}.  

Crucially, the operators which give rise to neutrino masses in this model may also give rise to
lepton flavour violation in the charged sector.  In this paper, we shall consider 
combinations of lepton number violating parameters that correctly reproduce the observations made in 
oscillation experiments.  For these sets of parameters we shall investigate whether
they would result in branching ratios of rare leptonic decays which would already have been observed, 
or would be observed by forthcoming experimental studies.  If the rare leptonic decays are not observed, 
the improving bound will be valuable in precluding certain scenarios.  We
select scenarios in which the off-diagonal terms in the supersymmetry breaking scalar mass matrices are
zero.  It is possible, of course, even in the R-parity conserving MSSM, that
this is not the case and that these terms will lead to large branching
ratios for lepton flavour violating decays~\cite{deCarlos:1995ah}.  The aim of this study is to
examine, specifically, the effects of lepton number violating terms in the
Lagrangian and the interplay between the charged and neutral sector.  As such,
we will examine the scenarios which are only present in the \LMSSM~and will not 
examine phenomena which have their origin in the R-parity conserving part of the Lagrangian. 

\section{Experimental Results, Bounds and Prospects}

The results from oscillation experiments combine to describe the mass squared
differences and mixing angles of the neutrino sector increasing accurately.
The current $3\sigma$ allowed ranges are~\cite{Schwetz:2006dh} 

\begin{eqnarray}
\sin^2\theta_{12} =0.24-0.40 \;, \qquad \sin^2\theta_{23} =0.34-0.68
\;, \qquad \sin^2\theta_{13} \le 0.041 \;, \label{nudata1} \\[3mm]
\Delta m_{21}^2 =(7.1-8.9)\times 10^{-5} \; {\rm eV}^2\;, \qquad
|\Delta m_{31}^2| =(1.9-3.2)\times 10^{-3}\; {\rm eV}^2\;.
\label{nudata2}
\end{eqnarray}
In our analysis we choose Lagrangian parameters such that the neutrino 
mixing angles match the tri-bimaximal mixing scenario of Ref.~\cite{Harrison:2002er},
\begin{eqnarray}
\sin^2\theta_{12} =\frac{1}{{3}} \;, \qquad 
\sin^2\theta_{23} =\frac{1}{{2}} \;, \qquad  \sin^2\theta_{13} = 0  \;.
\label{input}
\end{eqnarray}
The following bounds have been set on the branching ratios of $\mu
\rightarrow e \gamma$~\cite{Ahmed:2001eh}, $\tau \rightarrow \mu
\gamma$~\cite{Aubert:2005ye} and $\tau \rightarrow e
\gamma$~\cite{Aubert:2005wa}.

\begin{eqnarray}
	\textup{Br}(\mu \rightarrow e \gamma) < 1.2 \times 10^{-11}
	\hspace{30pt} \textup{at} \,\,  90\% \, \, \textup{CL} \label{br1} \\ 
	\textup{Br}(\tau \rightarrow \mu \gamma) < 6.8 \times 10^{-8}
	\hspace{30pt} \textup{at} \,\,90\% \, \, \textup{CL} \label{br2} \\ 
	\textup{Br}(\tau \rightarrow e \gamma) < 1.1 \times 10^{-7}
	\hspace{30pt} \textup{at} \,\,90\% \, \, \textup{CL} \label{br3}
\end{eqnarray}

Future experiments will probe these decays further.  It is
suggested~\cite{Mori,Giorgi,Calibbi:2006nq} that the sensitivity to $\mu \rightarrow e \gamma$
will be improved to $ \sim 10^{-13,-14}$, and that the sensitivities to $\tau \rightarrow
\mu \gamma$ and $\tau \rightarrow e \gamma$ will reach $\sim 10^{-8,-9}$.  To
conclude, we present the current experimental values for the branching ratios of 
$\tau \rightarrow \mu \nu_{\tau}
\bar{\nu}_{\mu}$, $\tau \rightarrow e \nu_{\tau} \bar{\nu}_{e}$ and
$\mu \rightarrow e \nu_{\mu} \bar{\nu}_{e}$~\cite{Yao:2006px},
\begin{eqnarray}
	\textup{Br}(\tau \rightarrow \mu \nu_{\tau} \bar{\nu}_{\mu}) =& 0.1736 \pm 0.0005
	\label{PDG1}\\ 
	\textup{Br}(\tau \rightarrow e \nu_{\tau} \bar{\nu}_{e}) =& 0.1784 \pm 0.0005  
	\label{PDG2}\\ 
	\textup{Br}(\mu \rightarrow e \nu_{\mu} \bar{\nu}_{e}) \approx&
	\hspace{-70pt} 1 \label{PDG3}
\end{eqnarray}

\section{Generic Diagrams for $l_i \rightarrow l_j \gamma$}
\label{sec:generic}

\begin{figure}
{\begin{center}
		\includegraphics[width=2in]{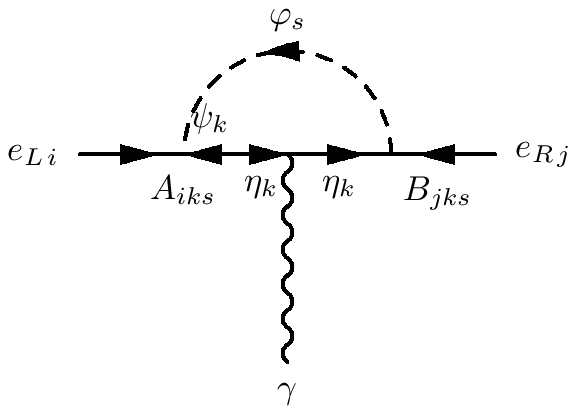} \hspace{10pt} 
\includegraphics[width=2in]{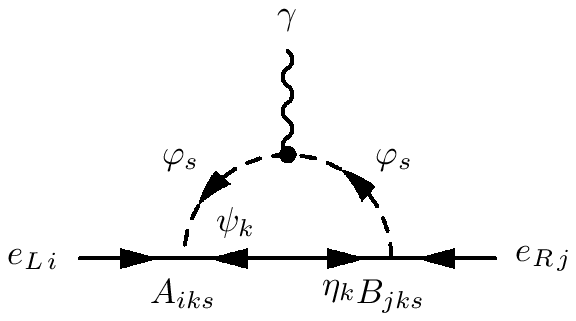}  \hspace{10pt}
\includegraphics[width=2in]{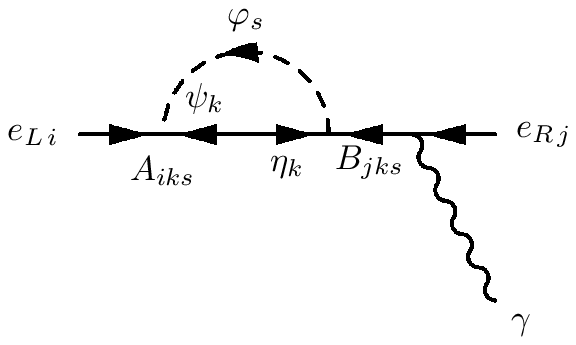}
\caption{\it The final state photon can be attached to either of the external fermions,
the fermion in the loop, or the scalar in the loop.  All combinations of
helicities of fermions should be included in the calculation.}
\label{dia:generic}
\end{center}}
\end{figure}

At the level of one loop, three basic types of diagram contribute
to the decay $l_i \rightarrow l_j \gamma$ and in each case, there is a fermion - boson loop.
The external photon can be attached either to the fermion in the loop, the 
boson in the loop or the external leg (Fig.~\ref{dia:generic}).  The
calculation, as shown in Fig.~\ref{dia:generic}, was performed in Weyl
notation.  In this notation, the four-component spinor $e \equiv \left( \begin{array}{c}
	e_L \\ \bar{e}_R \end{array} \right)$, where $e_L$ and $e_R$ are
	two-component left-handed spinors and the four-component
	spinor, $f \equiv \left( \begin{array}{c}
	\psi \\ \bar{\eta} \end{array} \right)$ denotes a generic fermion.
The factors associated with the vertices are denoted by either $A_{iks}$ or
$B_{iks}$ as shown in the diagrams, the charges of the particles are
given by $Q_{\psi,\varphi}$, the masses of the particles in the loop are 
$m_{{\psi,\varphi}}$ and the masses of the charged leptons on the external legs
are given by $m_{i,j}$.  
For more information concerning calculations using two-component spinors, see
Ref.~\cite{Habpresusy}.  Taking all possible combinations of arrows and 
neglecting diagrams with gauge bosons\footnote{
It can be seen that diagrams will be suppressed either by the magnitude of the
neutrino mass or, in diagrams which contain lepton number violating
operators, by the amount of mixing between the neutrinos/neutralinos and
charged leptons/charginos.}
it can be seen that, in agreement with Ref.~\cite{Brignole:2004ah,DHR}, at leading 
order the branching ratio is given by

\begin{equation}
\Gamma (l_i \rightarrow l_j \gamma) = \frac{48 \pi^2}{G^2_F m_i^2} \Big[
|\Lambda_L|^2 + |\Lambda_R|^2  \Big]
 \Gamma (l_i \rightarrow l_j  \nu_i \bar{\nu}_j) \; ,
\label{eq:br}
\end{equation}

\noindent where

$$\Lambda_R=\frac{1}{(4 \pi)^2} A_{iks} B_{jks} Q_{\psi_k} m_{\psi_k}
\left[ - \frac{ m^2_{\psi_k}-3m^2_{\phi_s} }{ 4(m^2_{\psi_k}-m^2_{\phi_s})^2  } 
+ \frac{m^4_{\phi_s}}{(m^2_{\psi_k}-m^2_{\phi_s})^3} 
\ln \left[ \frac{m_{\phi_s}}{m_{\psi_k}} \right] \right]$$
$$ +\frac{1}{2(4 \pi)^2} A_{iks} B_{jks} Q_{\phi_s} m_{\psi_k}
\left[ - \frac{m^2_{\psi_k}+m^2_{\phi_s}}{2(m^2_{\psi_k}-m^2_{\phi_s})^2} 
+ \frac{2m^2_{\psi_k}m^2_{\phi_s}}{(m^2_{\psi_k}-m^2_{\phi_s})^3} 
\ln \left[ \frac{m_{\phi_s}}{m_{\psi_k}} \right] \right]$$
$$-\frac{1}{2(4 \pi)^2} B^*_{iks} B_{jks} Q_{\psi_k} m_{i}
\left[- \frac{m^4_{\psi_k}-5m^2_{\psi_k}m^2_{\phi_s}-2m^4_{\phi_s}}{12(m^2_{\psi_k}-m^2_{\phi_s})^3} 
+ \frac{m^2_{\psi_k}m^4_{\phi_s}}{(m^2_{\psi_k}-m^2_{\phi_s})^4} 
\ln \left[ \frac{m_{\phi_s}}{m_{\psi_k}} \right] \right]$$
$$-\frac{1}{2(4 \pi)^2} B^*_{iks} B_{jks} Q_{\phi_s} m_{i}
\left[ \frac{m^4_{\phi_s}-5m^2_{\psi_k}m^2_{\phi_s}-2m^4_{\psi_k}}{12(m^2_{\psi_k}-m^2_{\phi_s})^3} 
- \frac{m^2_{\phi_s}m^4_{\psi_k}}{(m^2_{\psi_k}-m^2_{\phi_s})^4} 
\ln \left[ \frac{m_{\phi_s}}{m_{\psi_k}} \right] \right]$$
and
$$\Lambda_L=\frac{1}{(4 \pi)^2} A^*_{jks} B^*_{iks} Q_{\psi_k} m_{\psi_k}
\left[ - \frac{ m^2_{\psi_k}-3m^2_{\phi_s} }{ 4(m^2_{\psi_k}-m^2_{\phi_s})^2  } 
+ \frac{m^4_{\phi_s}}{(m^2_{\psi_k}-m^2_{\phi_s})^3} 
\ln \left[ \frac{m_{\phi_s}}{m_{\psi_k}} \right] \right]$$
$$ +\frac{1}{2(4 \pi)^2} A^*_{jks} B^*_{iks} Q_{\phi_s} m_{\psi_k}
\left[ - \frac{m^2_{\psi_k}+m^2_{\phi_s}}{2(m^2_{\psi_k}-m^2_{\phi_s})^2} 
+ \frac{2m^2_{\psi_k}m^2_{\phi_s}}{(m^2_{\psi_k}-m^2_{\phi_s})^3} 
\ln \left[ \frac{m_{\phi_s}}{m_{\psi_k}} \right] \right]$$
$$-\frac{1}{2(4 \pi)^2} A^*_{jks} A_{iks} Q_{\psi_k} m_{i}
\left[- \frac{m^4_{\psi_k}-5m^2_{\psi_k}m^2_{\phi_s}-2m^4_{\phi_s}}{12(m^2_{\psi_k}-m^2_{\phi_s})^3} 
+ \frac{m^2_{\psi_k}m^4_{\phi_s}}{(m^2_{\psi_k}-m^2_{\phi_s})^4} 
\ln \left[ \frac{m_{\phi_s}}{m_{\psi_k}} \right] \right]$$
$$-\frac{1}{2(4 \pi)^2} A^*_{jks} A_{iks} Q_{\phi_s} m_{i}
\left[ \frac{m^4_{\phi_s}-5m^2_{\psi_k}m^2_{\phi_s}-2m^4_{\psi_k}}{12(m^2_{\psi_k}-m^2_{\phi_s})^3} 
- \frac{m^2_{\phi_s}m^4_{\psi_k}}{(m^2_{\psi_k}-m^2_{\phi_s})^4} 
\ln \left[ \frac{m_{\phi_s}}{m_{\psi_k}} \right] \right]
\; .
$$

The contribution from each Feynman diagram was expanded under the assumption
$m_i^2,m_j^2 \ll m_\varphi^2, m_\psi^2$.  On-shell conditions could then be applied and
terms proportional to $m_j$ were neglected.  The resulting expression can then be re-arranged
to be seen to contribute to effective operators of the form\footnote{
We define $\sigma^{\mu \nu} \equiv \frac{1}{4} (\sigma^\mu \bar{\sigma}^\nu
- \sigma^\nu \bar{\sigma}^\mu)$ and
$\bar{\sigma}^{\mu \nu} \equiv \frac{1}{4} (\bar{\sigma}^\mu \sigma^\nu
- \bar{\sigma}^\nu \sigma^\mu)$}

\begin{equation}
	{\cal L}_{\textup{\tiny{eff}}} \supset 2 \Lambda_{R\,ij} e_{R\,j} \sigma^{\mu \nu} e_{L\,i} F_{\mu \nu}
	+ 2 \Lambda_{L\,ij} \bar{e}_{L\,j} \bar{\sigma}^{\mu \nu}
	\bar{e}_{R\,i} F_{\mu \nu} \;.
\end{equation}
Individually, diagrams produce terms contributing to different effective
operators, but these cancel when all possible diagrams are
considered.  Following Ref.~\cite{Brignole:2004ah} 
and inserting the experimental values given in Eqs.~(\ref{PDG1},\ref{PDG2},\ref{PDG3}), 
it is then possible to
move from this effective operator to a branching ratio for the rare decay,
given by Eq.~(\ref{eq:br}).  

\section{Specific Diagrams for $l_i \rightarrow l_j \gamma$}
\label{sec:vertices}
{
	\begin{figure}
\begin{center}
		\includegraphics[width=2.2in]{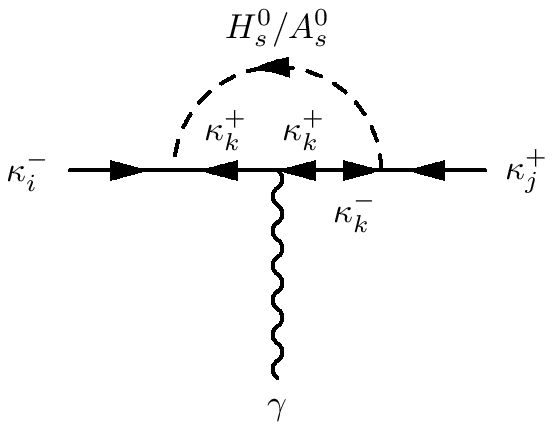} \hspace{20pt}
\includegraphics[width=2.2in]{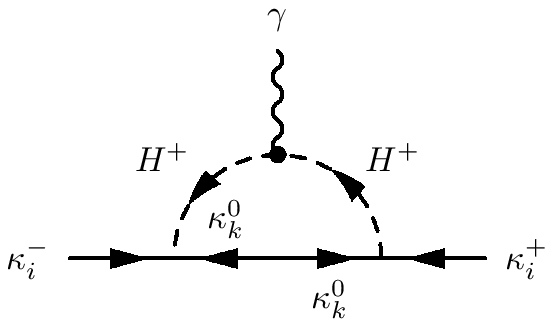} \\ 
\includegraphics[width=2.2in]{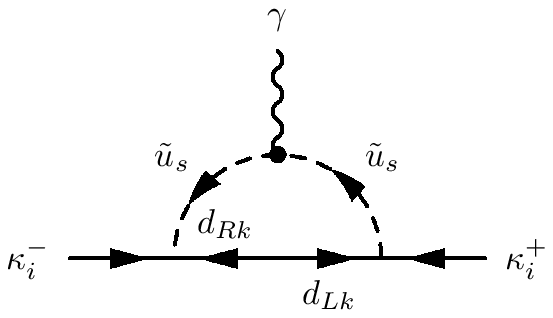} \hspace{20pt}
\includegraphics[width=2.2in]{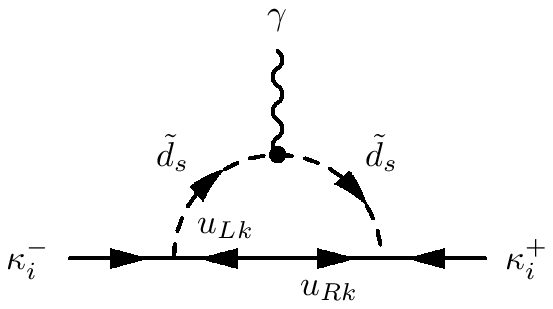}
\caption{\it The various possible combinations of particles which can be produced
in the loop: chargino and neutral scalar; neutralino and charged scalar; 
down quark and up-type squark; up quark and down-type squark.}
\label{dia:spec}
\end{center}
\end{figure}
}

The following combinations of particles can be produced inside the loop:
chargino and neutral scalar; neutralino and charged scalar; quark and squark,
as shown in Fig.~\ref{dia:spec}.  In the \LMSSM, mixing occurs between
charged leptons/charginos and between neutrinos/neutralinos.  For example,
there are five charged fermions,
$$\chi_i^{\pm} = \left( \begin{array}{c}
	\kappa_i^{\pm} \\ \bar{\kappa}_i^{\mp} \end{array} \right) \;, \;
	i=1,\ldots,5 \;,$$
	where $i=3,4,5$ are the charged leptons $e$, $\mu$ and $\tau$.
	Similarly, there are seven neutral fermions,
$$\chi_i^{0} = \left( \begin{array}{c}
\kappa_i^{0} \\ \bar{\kappa}_i^{0} \end{array} \right) \;, \;
i=1,\ldots,7 \;,$$
where $i=5,6,7$ are the neutrinos.  The two-component spinors comprising the
quarks are denoted,
$$d_i = \left( \begin{array}{c}
	d_{L\;i} \\ \bar{d}_{R\,i} \end{array} \right)\;, \;
i=1,\ldots,3 \;,$$
$$u_i = \left( \begin{array}{c}
	u_{L\;i} \\ \bar{u}_{R\,i} \end{array} \right)\;, \;
i=1,\ldots,3 \;.$$

Each of the diagrams in Fig.~\ref{dia:spec} are in the same form as
outlined in Sec.~\ref{sec:generic}.  The generic vertices $A_{iks}$ and
$B_{jks}$ can be replaced by the appropriate Feynman rule, which are presented
in the Appendix.  The calculation is performed in the mass eigenbasis.  The
full mass matrices are diagonalised and the appropriate rotation matrices are
calculated numerically and without approximation.  In understanding the important
physical contributions it is more useful, however,
to present diagrams in the mass insertion approximation containing interaction
state particles.  The
plots are based on a Fortran code which computes the full result.
 
We will consider the role played by combinations of lepton number violating
parameters by, first, investigating the case in which the
bilinear lepton number violating coupling in the superpotential
correctly produces the atmospheric mass
difference (and the ratios between the three components ensure the mixing angles are 
reproduced correctly) and another, single, lepton number violating coupling sets the solar mass
difference.  Both sources of lepton number violation will then combine to
produce a diagram which contributes to a lepton flavour violating decay.  
Second, we will consider the scenario in which both the scale of the atmospheric 
mass difference and the solar scale are set by radiative corrections, and the
bilinear lepton number violating parameters are set to zero.

\subsection{Atmospheric scale set by \boldmath$\mu_{1,2,3}$}

\begin{figure}
{\begin{center}
	\includegraphics{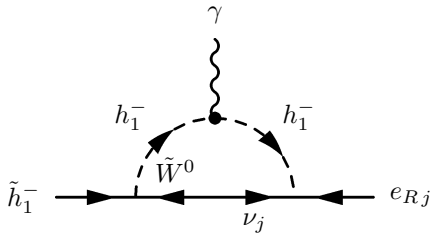}   
\caption{\it Diagram contributing to $l \rightarrow l' \gamma$ with the only
source of lepton number violating being the bilinear lepton number violating
couplings in the superpotential, $\mu_i$.}
\label{dia:mu}
\end{center}}
\end{figure}

\begin{figure}
{\begin{center}
	\includegraphics[width=3.25in]{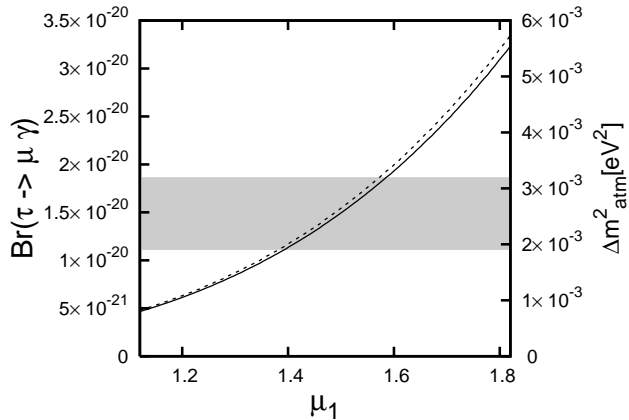}   
\end{center}}
\caption{\it $\tau \rightarrow \mu \gamma$ and $\Delta m^2_{atm}$ against $\mu_{1}$
-- the dotted line (and right hand axis) show the magnitude of $\Delta
m^2_{\textup{\tiny{atm}}}$ with the light grey band indicating the current $3\sigma$
allowed range; 
the full line (and left hand axis) show the branching
ratio of $\tau \rightarrow \mu \gamma$.}
\label{plot:atmtau}
\end{figure}

With only $\mu_{1,2,3} \neq 0$ and all other lepton number violating couplings
set to zero the atmospheric mass squared difference can be correctly
reproduced; the solar mass squared difference is not generated and no
observable branching ratios for $l
\rightarrow l' \gamma$ are generated. 
The non-zero $\mu_{1,2,3}$ do bring about a branching ratio for $l \rightarrow l' \gamma$, 
through the diagram shown in Fig.~\ref{dia:mu}.  The
fermion inside the loop is a mixture of the heavy neutralinos and the
interaction state neutrinos.  The amount of mixing between these interaction
states is dependent on $\mu_{1,2,3}$, and also 
determines the mass of the tree level neutrino.  The amount of
mixing between the external leg interaction state charged higgsino and left handed charged leptons,
is also determined by $\mu_{1,2,3}$.  As such,
this diagram contributes to the $l_i \rightarrow l_j \gamma$ decay with
branching ratio approximately given by,

\begin{equation}
	\Gamma(l_i \rightarrow l_j \gamma) \approx
\frac{3}{(4\pi)^2} \frac{|\lambda_{0jj}|^2}{G_F^2 m_i^2} 
 \frac{e^2}{\sw^2} M^2_{\chi^0}
\left[ 
\left( \frac{1}{M^2_{H^-}}\right)
\left(\frac{\mu_{0} \mu_{i}}{M_{\chi^{\pm}}^2} \right)
\left(\frac{\mu_{j} g_2 v_u}{M^2_{\chi^0}} \right) 
\right]^2
\Gamma (l_i \rightarrow l_j  \nu_i \bar{\nu}_j) \; .
\end{equation}

We will consider this scenario.  The $\mu_{1,2,3}$ parameter takes the values
\begin{equation}
	\mu_1 = \frac{\mu_2}{\sqrt{2}} = \frac{\mu_3}{\sqrt{3}} =
	1.12 -  1.82 \textup{MeV} \; ,
\end{equation}
\noindent where ratios between components of $\mu_i$ are chosen such that the mixing angles
in the PMNS matrix are generated to take the tri-bimaximal form, which are in
agreement with current bounds, and are entirely generated in the neutral sector.
A more detailed presentation of all the following scenarios is presented in Ref.~\cite{Dedes:2006ni}.
The overall scale is varied and the resulting mass squared difference and branching ratio
for $\tau \rightarrow \mu \gamma$ are calculated and shown in Fig.~{\ref{plot:atmtau}.  
The light grey band (and right hand axis) shows the current $3\sigma$ allow
region for the atmospheric mass squared difference as given by
Eq.~(\ref{nudata2}).     
All other lepton number violating couplings are set to
zero, and R-parity conserving parameters are fixed to be the SPS1a benchmark
point~\cite{Allanach:2002nj}.  We note that, in agreement with Ref.~\cite{Carvalho:2002bq}, the branching ratios for $l_i
\rightarrow l_j \gamma$ are well below current experimental limits, as given in Eq.~(\ref{br2}), and show the
resulting branching ratio for $\tau \rightarrow \mu \gamma$ in Fig.~{\ref{plot:atmtau}.

\subsection{Atmospheric scale set by \boldmath$\mu_{1,2,3}$ -- Solar scale set by $\lambda_{ikk}$}

\begin{figure} 
{\begin{center}
	\includegraphics{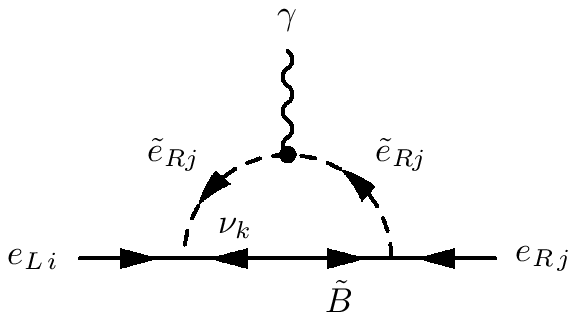} \hspace{30pt} 
\includegraphics{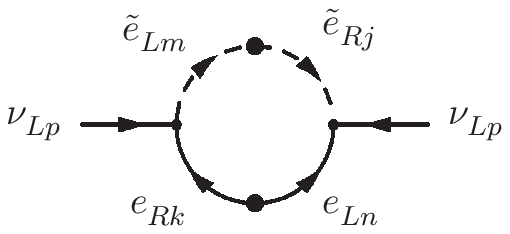} 
\caption{\it Feynman diagrams generated when, firstly, $\mu_i$
induces mixing of interaction state neutrinos with gauginos and
higgsinos and secondly, $\lambda_{ijk} \neq 0$ generating neutrino masses
radiatively.}
\label{dia:kaplam}
\end{center}}
\end{figure}

For the remaining examples, the bilinear lepton number couplings take the
values,
\begin{equation}
	\mu_1 = \frac{\mu_2}{\sqrt{2}} = \frac{\mu_3}{\sqrt{3}} =
	1.47 \textup{MeV} \; ,
\end{equation}
which reproduce correctly the atmospheric mass squared difference, as
shown in Fig.~\ref{plot:atmtau}.
A single, further lepton number violating coupling 
$\lambda_{ikk}(=-\lambda_{kik})$ is then varied.  The
branching ratio for lepton flavour violating decays when 
this coupling correctly generates the observed solar flavour
oscillation are calculated numerically using a 
Fortran code.\footnote{
The same code was used in Ref.~\cite{Dedes:2006ni} to calculate
the neutrino masses.}  We find that there are scenarios of this form which correctly
reproduce all neutrino data and give rise to branching ratios for $l
\rightarrow l' \gamma$ which are, or will be, observable in experimental
studies.

When $\mu_{1,2,3},\lambda_{ikk} \neq 0$ then diagrams shown in 
Fig.~\ref{dia:kaplam} are generated.  It can be seen that the 
amount of mixing on the fermion line
inside the loop, again, corresponds directly to the amount of mixing between interaction
state neutrinos and gauginos/higgsinos.  It is this mixing
which determines the mass of the neutrino produced at tree level and is
determined by the values given to the bilinear lepton number violating
parameters, $\mu_i$.   The left hand vertex is determined by the $\lambda$ 
coupling from the superpotential, as defined in Eq.~(\ref{superpot1}).  This term in the superpotential generates
both couplings in the second diagram of Fig.~\ref{dia:kaplam}.  In fact, if
$\lambda_{ikk} \neq 0$, that is, any $\lambda$ coupling with the final two
indices the same, a single $\lambda$ coupling will generate this diagram.   
The branching ratio is approximately given by the following expression,
\begin{equation}
	\Gamma(l_i \rightarrow l_j \gamma) \approx
	\frac{3}{(4\pi)^2} \frac{|\lambda_{ikk}|^2}{G_F^2 m_i^2} 
	 \, \frac{e^2}{\cw^2} \, M^2_{\chi^0}
	\left[ 
\left(\frac{1}{m_{\tilde{e}}^2} \right) \, 
\left(\frac{\mu_k g v_u}{M^2_{\chi^0}} \right) \right]^2
\Gamma (l_i \rightarrow l_j  \nu_i \bar{\nu}_j) \; .
\label{eq:brlambda}
\end{equation}
In the following sections, we shall consider in turn all $\lambda$ couplings with symmetric final indices.

\subsubsection{\boldmath$\mu_{1,2,3}$ and $\lambda_{211}$}
\label{sec:lam121}

\begin{figure}
{\begin{center}
	\includegraphics[width=3.25in]{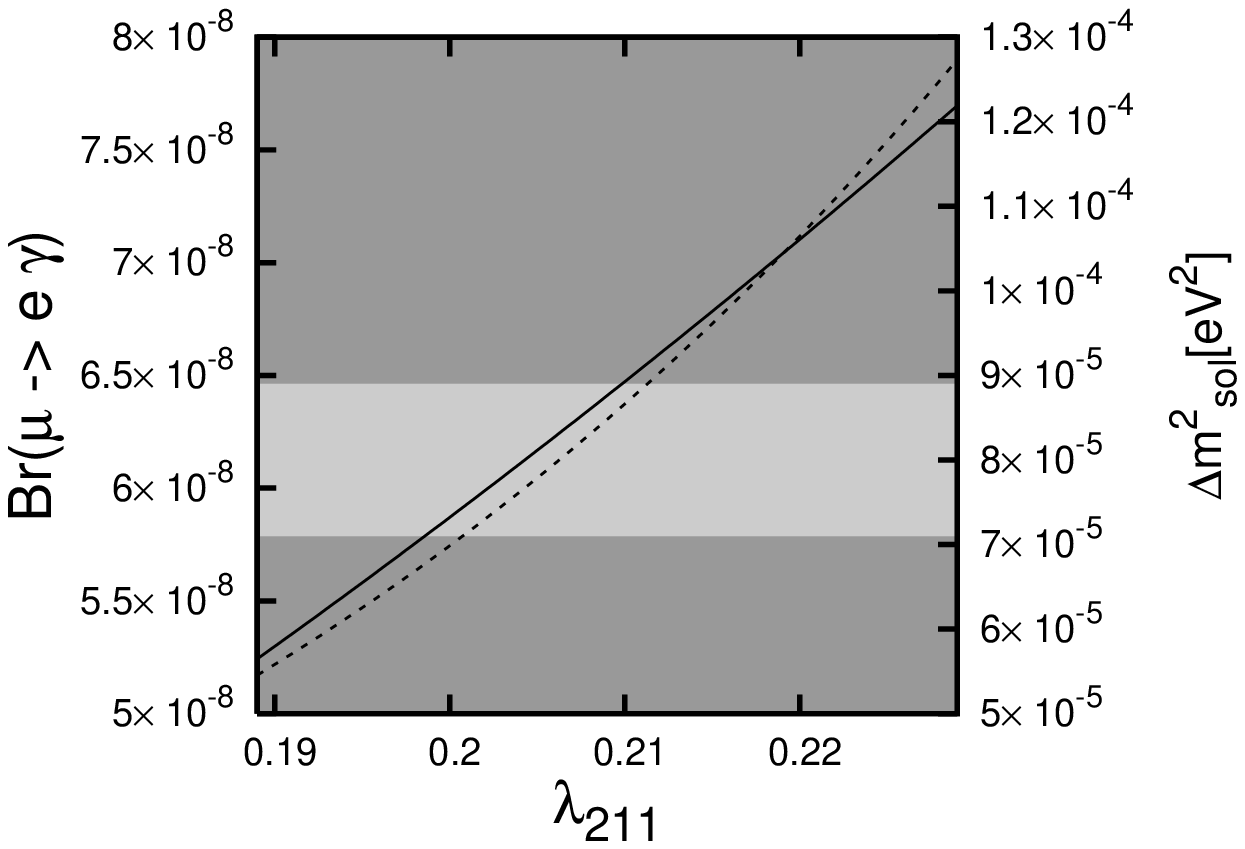}   
	\includegraphics[width=3.25in]{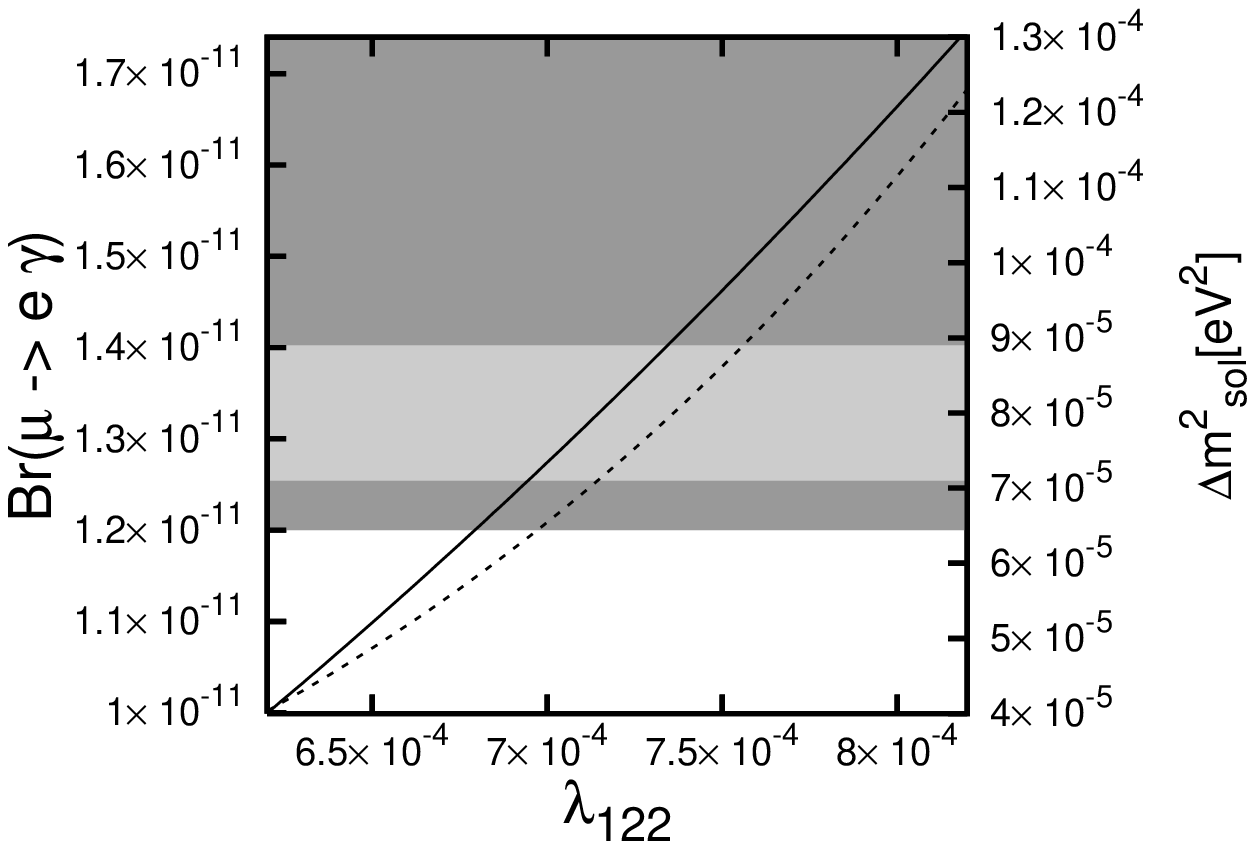}
	\includegraphics[width=3.25in]{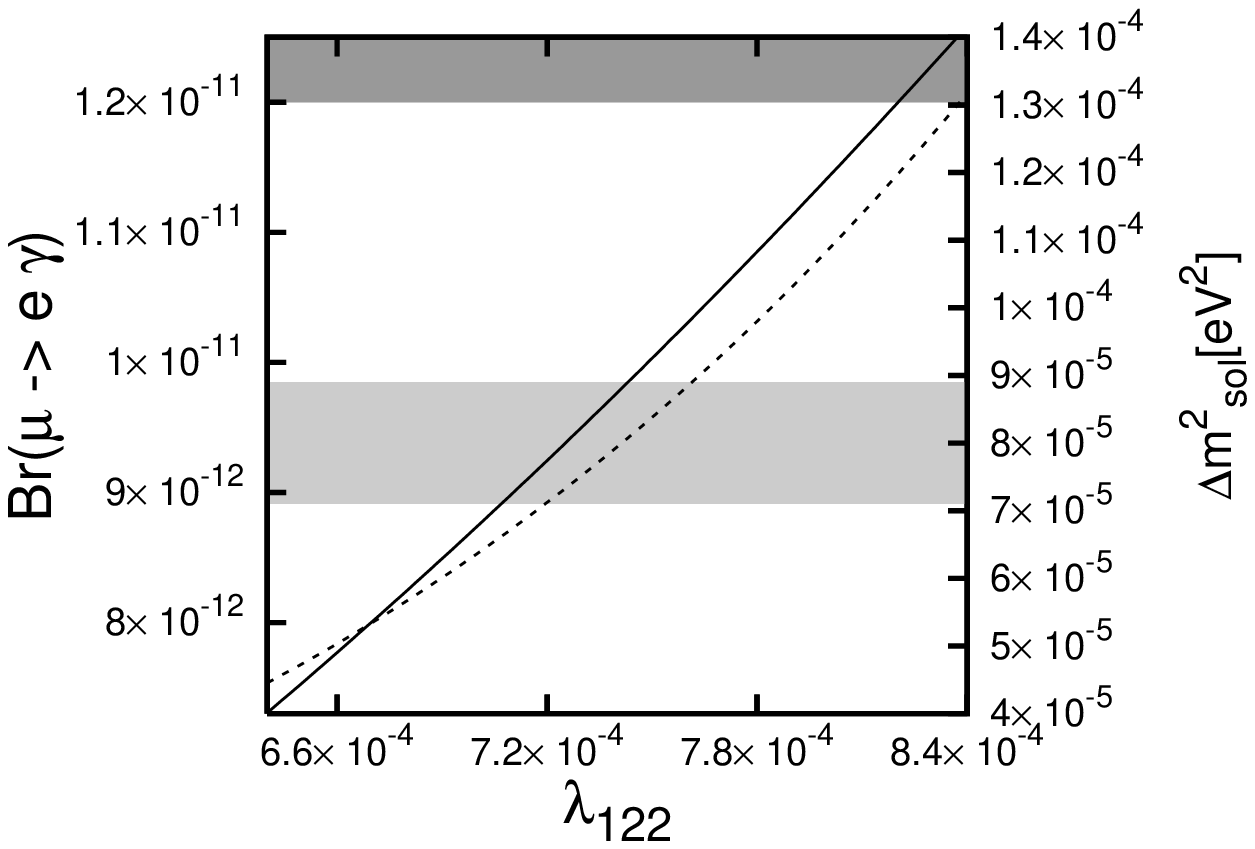}
	\includegraphics[width=3.25in]{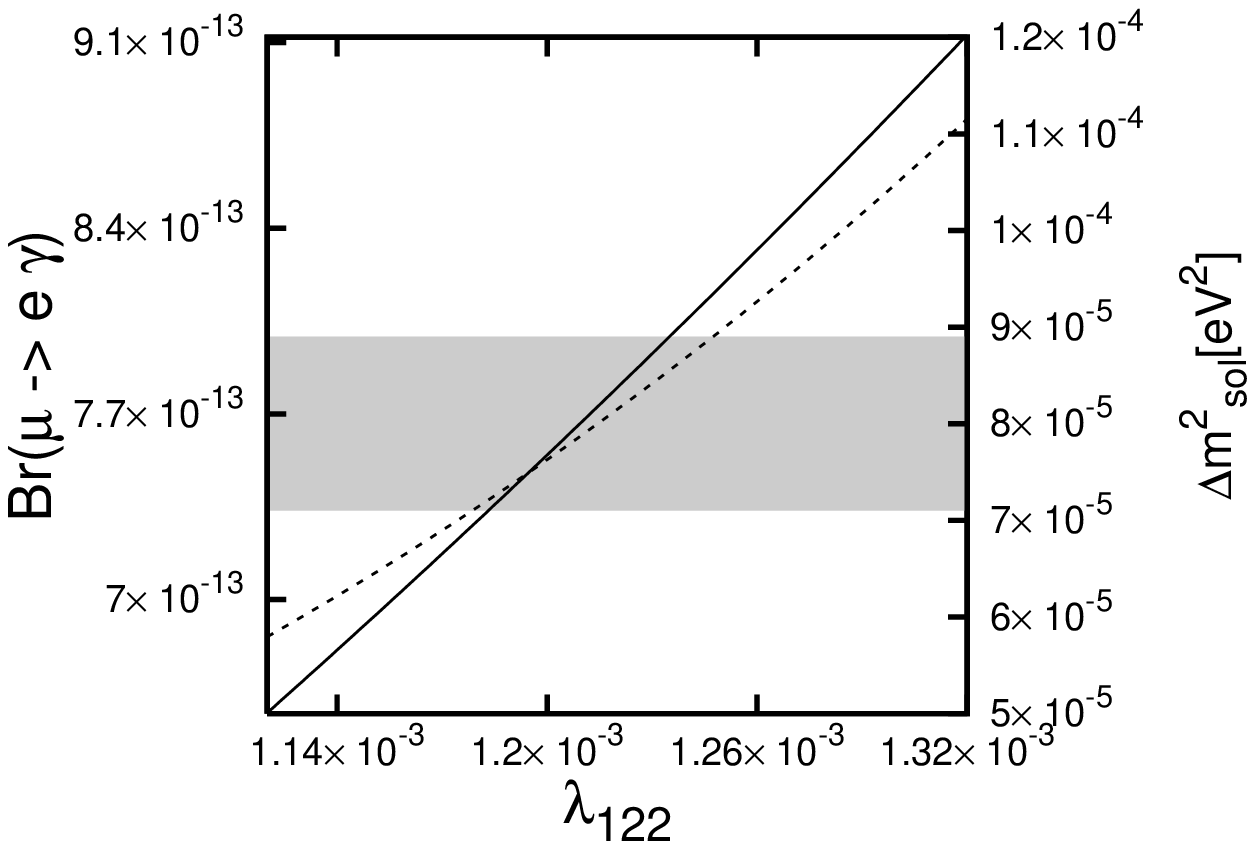}
\end{center}}
\caption{\it $\mu \rightarrow e \gamma$ and $\Delta m^2_{sol}$ against $\lambda_{211}$ or $\lambda_{122}$
-- the dotted line (and right hand axis) show the magnitude of $\Delta
m^2_{\textup{\tiny{sol}}}$ with the light grey band indicating the current $3\sigma$
allowed range; 
the full line (and left hand axis) show the branching
ratio of $\mu \rightarrow e \gamma$ with the dark grey area showing the values
currently excluded at 90\% confidence level.  For the upper two plots, the
R-parity conserving parameters are set by the SPS1a benchmark points, for
which the mass of the charged scalar that consists mostly of interaction state
$\tilde{\mu}_R$ is approximately 143 GeV.  For the lower two plots, the mass of
the scalars was increased such that the mass of $\tilde{\mu}_R$ was raised to
145 GeV (bottom left) and 265 GeV (bottom right).
}
\label{plot:lam121mu}
\end{figure}

The first example considered is $\lambda_{211}$.  As $\lambda_{211}$ is varied,
Fig.~\ref{plot:lam121mu} shows the resulting solar mass squared difference,
given by the dashed line, and the $\mu \rightarrow e \gamma$ branching ratio,
given by the full line.  The light grey strip
shows the current experimental value for the solar mass squared difference and
the dark grey area is the area presently excluded by $\mu \rightarrow e
\gamma$ searches.

The value of $\lambda_{211}$ required to generate the correct value for the
solar mass difference must be comparatively large to compensate for the
smallness of the mass of the electron induced in the loop.  The $\mu
\rightarrow e \gamma$ diagram generated has a large mass in the loop, and
there is no suppression in the slepton part of the graph due to
intergenerational, or left-right, mixing.  As such, it can be seen that this scenario,
although correctly explaining neutrino masses, is ruled out 
as it predicts a $\mu \rightarrow e \gamma$ branching ratio which would have
been observed already.  Furthermore, it is shown in Ref.~\cite{Barger:1989rk,Allanach:1999ic}, 
that a $\lambda_{211}$ coupling of this magnitude would violate charged current universality.

\subsubsection{\boldmath$\mu_{1,2,3}$ and $\lambda_{122}$}

In the top right plot of Fig.~\ref{plot:lam121mu}, it is shown that 
the value of $\lambda_{122}$ required to correctly generate the
solar mass squared difference is smaller; a muon is now produced in the
loop contributing to the neutrino mass.  The lower value of $\lambda_{122}$, 
in turn, makes the
$\mu \rightarrow e \gamma$ branching ratio produced lower than the previous example, 
however it is still at the edge of the region ruled out by experiment with 90\% confidence level.
For scenarios with slightly heavier scalar masses than the SPS1a
benchmark point this would not be ruled out.  In the two bottom plots of
Fig.~\ref{plot:lam121mu} the mass of the scalars has been increased, such that
the mass of the charged scalar which is mostly $\tilde{\mu}_R$ is 
143 GeV (bottom left) and 265 GeV (bottom right) compared to approximately
145 GeV which is produced by the SPS1a benchmark values of R-parity
conserving parameters.  We note how sensitive the resulting branching ratio is
to the mass of the scalar in the loop.  As shown in Eq.~(\ref{eq:brlambda}), the
$ \textup{Br}(l \rightarrow l' \gamma) \sim 1/m^4_{\tilde{\mu}}$.  Because
this scenario is at the edge of current limits, and because of this sensitivity to the value
of scalar masses, it is a particularly interesting scenario which can be studied
in future experiments.

\subsubsection{\boldmath$\mu_{1,2,3}$ and $\lambda_{311,133}$}

\begin{figure}
{\begin{center}
	\includegraphics[width=3.25in]{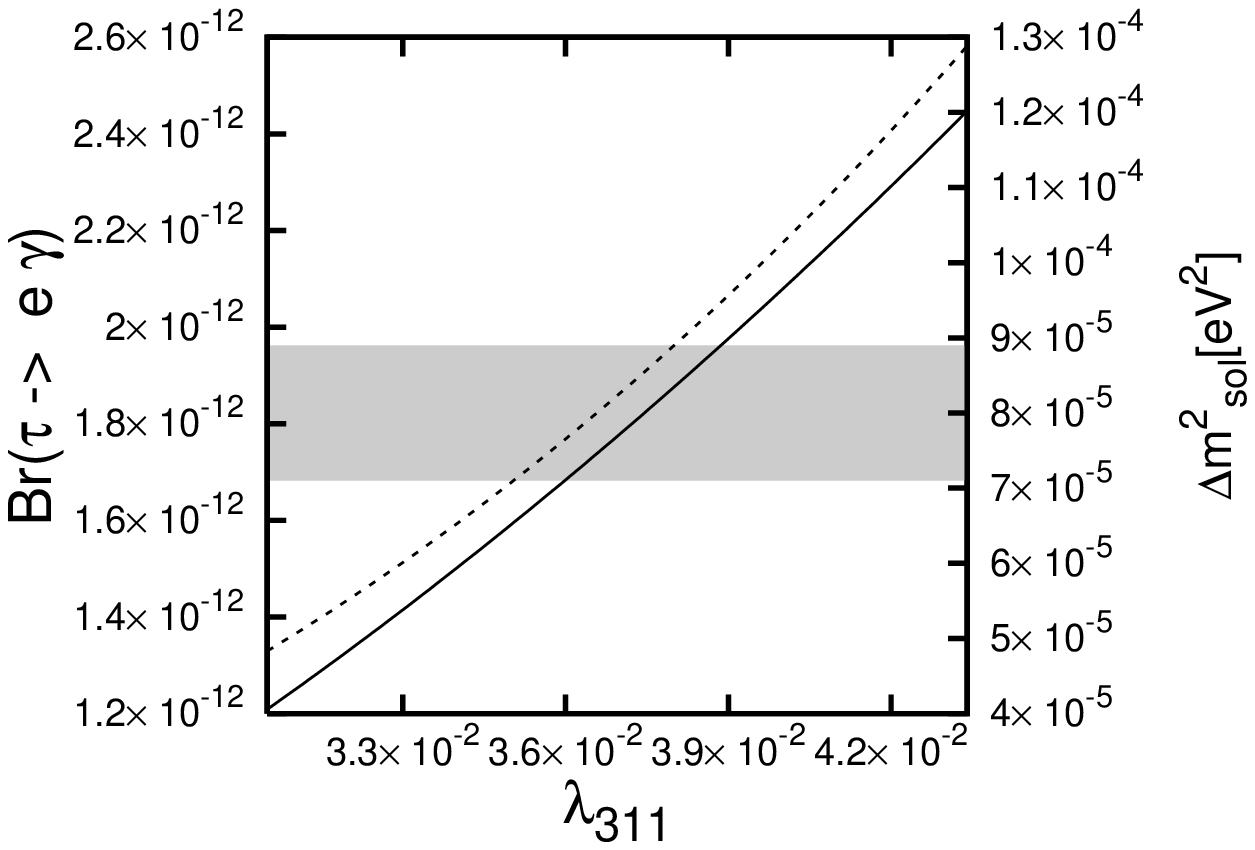} 
	\includegraphics[width=3.25in]{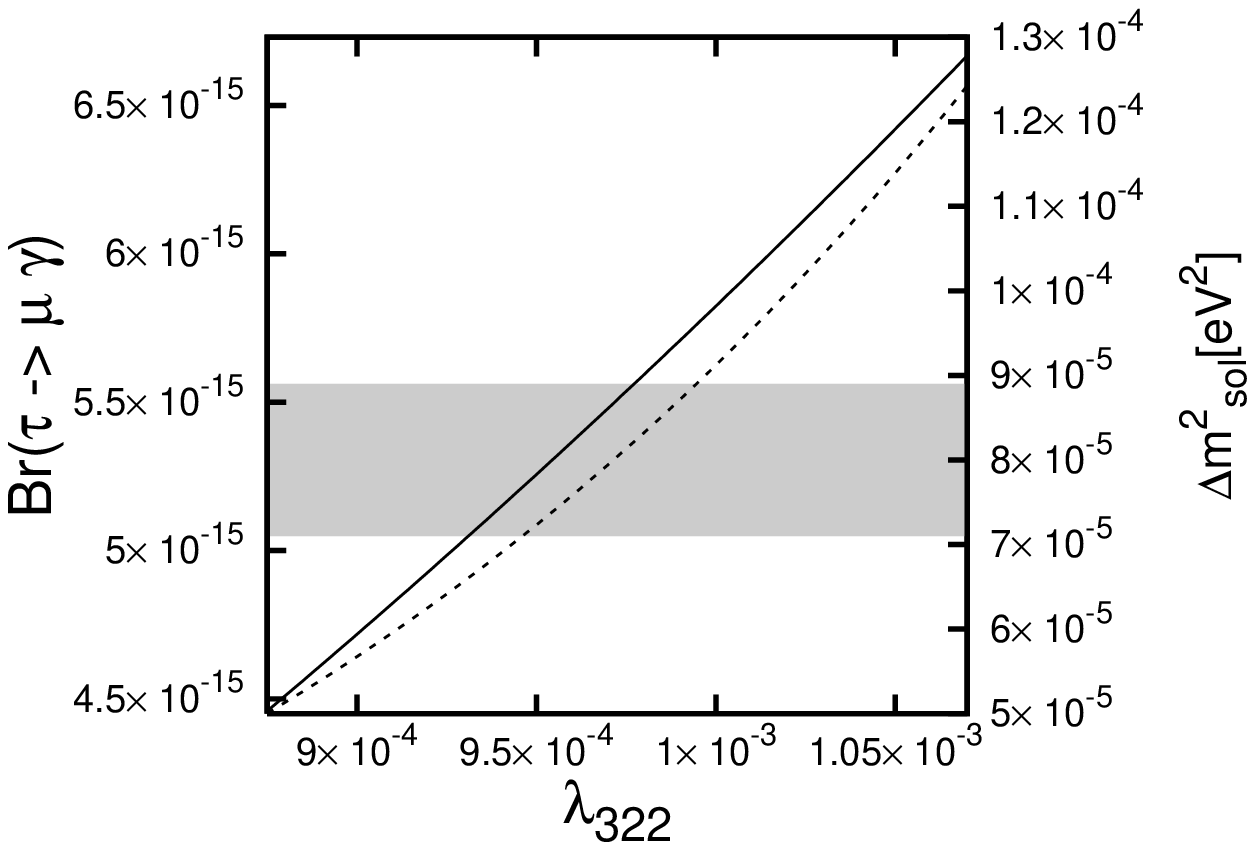}\\
	\includegraphics[width=3.25in]{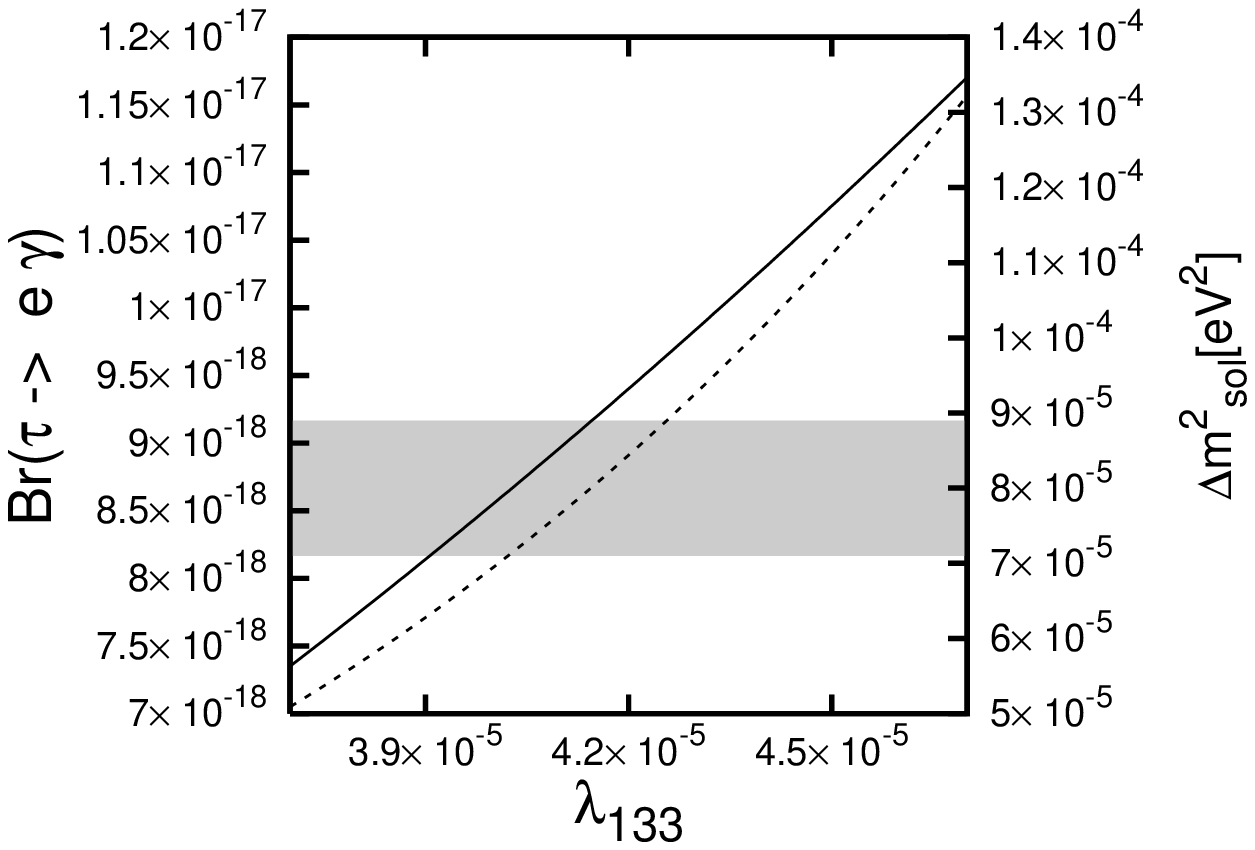}
	\includegraphics[width=3.25in]{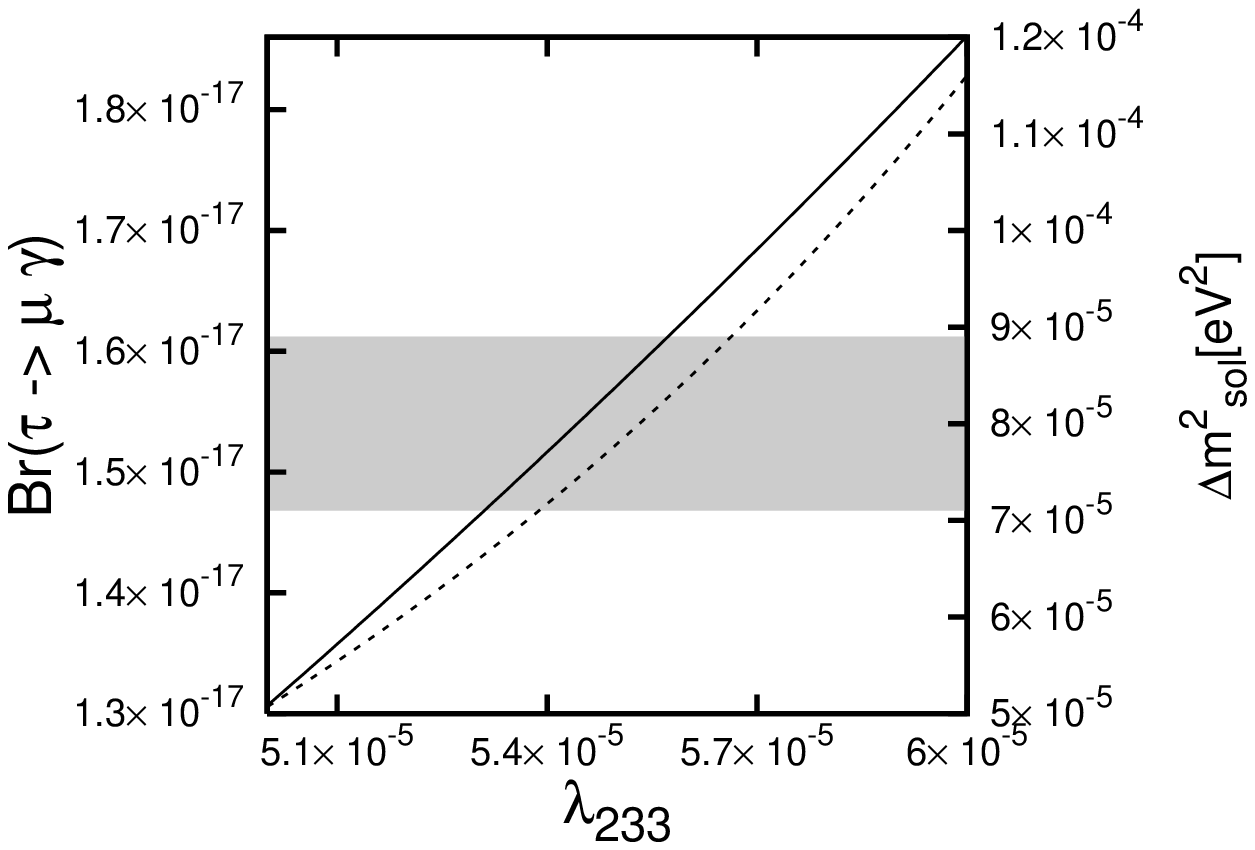}
\end{center}}
\caption{\it
$\tau \rightarrow e \gamma$ (full line, left axis) and $\Delta m^2_{sol}$ 
(dotted line, right axis) against $\lambda_{311,133}$ or $\lambda_{133}$;
$\tau \rightarrow \mu \gamma$ (full line, left axis) and $\Delta m^2_{sol}$ 
(dotted line, right axis) against $\lambda_{322}$ or $\lambda_{233}$
}
\label{plot:lam131te}
\end{figure}

In the first plot of Fig.~\ref{plot:lam131te}, we note the effect on the branching ratio of
$\tau \rightarrow e \gamma$ by varying $\lambda_{311}$.  The values for $\lambda_{311}$ which correctly
reproduce the neutrino data are not ruled out by current rare decay searches.
Not only is the experimental bound less stringent, but the branching ratio is suppressed by a factor of
$(m_\tau^2/m_\mu^2)(\textup{Br}(\mu \rightarrow e \nu_{\mu} \bar{\nu}_{e})/
\textup{Br}(\tau \rightarrow e \nu_{\tau} \bar{\nu}_{e})) \sim 1600$ compared 
to the that of Sect.~\ref{sec:lam121} and the fact
that the $\lambda$ coupling itself is smaller.  The predicted branching ratio generated by the
$\lambda_{133}$ coupling that correctly generates the solar mass squared difference, is even smaller due
to the lower value of the coupling.

\subsubsection{\boldmath$\mu_{1,2,3}$ and $\lambda_{322,233}$}

The right hand plots of Fig.~\ref{plot:lam131te} demonstrate that the values of
$\lambda_{322}$ or $\lambda_{233}$ which reproduce the neutrino results are not excluded and
well below current experimental sensitivity.  We
note that the values for $\lambda_{322}$ which produce the neutrino mass are
smaller than for $\lambda_{211}$ due to mass of the $\mu$
produced in the loop and $\lambda_{233}$ smaller still.

\subsection{Atmospheric scale set by \boldmath$\mu_{1,2,3}$ -- Solar scale set by  $\lambda'_{ikk}$}

\begin{figure}
\begin{center} \includegraphics{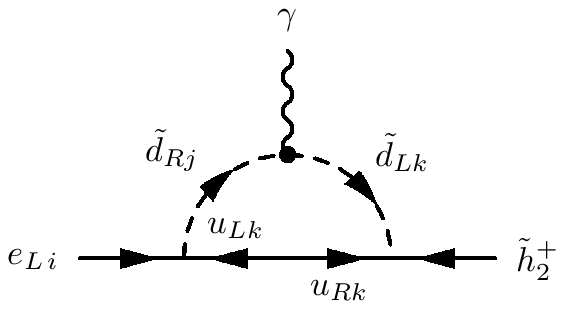} 
	\includegraphics{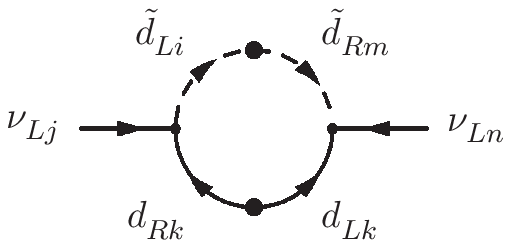}
	\caption{\it Feynman diagrams generated when $\lambda'_{ikk}$ generates
	neutrino masses radiatively and $\mu_{1,2,3}$ induce mixing on the
	external leg.}
	\label{dia:kaplamp}
\end{center}
\end{figure}

When $\mu_i,\lambda'_{ikk} \neq 0$ the diagrams shown in
Fig.~\ref{dia:kaplamp} are generated. In this case, 
the mixing on the external leg is driven by the $\mu_i$ term, 
again being determined such that the atmospheric mass difference is 
produced correctly at tree level.  In a similar fashion to the previous
section, the $\lambda'$ coupling on the left hand vertex is varied and 
the resulting solar mass squared difference considered.
The branching ratio is approximately given by
\begin{equation}
	\Gamma(l_i \rightarrow l_j \gamma) \approx
	\frac{3}{(4\pi)^2} \frac{|\lambda'_{ikk}|^2}{G_F^2 m_i^2} 
	\, |(Y_U)_{kk}|^2 \, m^2_{u_k}
	\left[ 3 
\left(	\frac{1}{m_{\tilde{d}}^2} \right) 
\left(\frac{\mu_k m_{e_k}}{m^2_{\chi^\pm}} \right) 
	\,\left( \frac{ {\cal M}^2_{\tilde{d}\,LR}}{ {\cal M}^2_{\tilde{d}\,R}- 
	{\cal M}^2_{\tilde{d}\,L}}\right)  \right]^2
\Gamma (l_i \rightarrow l_j  \nu_i \bar{\nu}_j) \; ,
\end{equation}
where ${\cal M}^2_{\tilde{d}\,L}$ and ${\cal M}^2_{\tilde{d}\,R}$ are diagonal
entries in the squark mass matrices and ${\cal M}^2_{\tilde{d}\,LR}$ is the
off-diagonal term which determines the mixing between the scalar partners of
the left and right handed quarks.
We note that the mixing between $e_R$ and charginos is much smaller than the
mixing of $\nu$ and neutralinos, and as such there is a suppression relative
to the $\lambda$-driven diagrams.  Furthermore, at the SPS1a benchmark point,
the squarks are heavier than the charged sleptons; the branching ratios are
highly sensitive to the scalar mass and this further suppresses $\lambda'$
contributions in comparison with $\lambda$ diagrams.
The result being that the $\lambda'$ vertices produce a negligible effect in this scenario.  

\begin{figure}
{\begin{center}
	\includegraphics[width=3.25in]{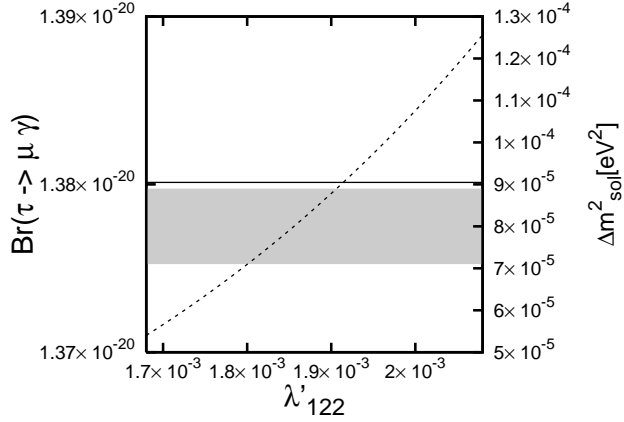} 
\end{center}}
\caption{\it $\tau \rightarrow \mu \gamma$ and $\Delta m^2_{sol}$ against $\lambda'_{122}$
.}
	\label{plot:lamp122tau}  
\end{figure}

With only the $\mu_i \neq 0$, setting the atmospheric mass
scale, the diagram shown in Fig.~\ref{dia:mu} gives a contribution to the $\tau \rightarrow \mu
\gamma$ branching ratio of the order $1.38\times 10^{-20}$ (Fig.~\ref{plot:atmtau}) 
compared to which we can 
ignore the contribution from quark loops, as shown in Fig.~\ref{plot:lamp122tau} 
which takes $\lambda'_{122}$ as an example.

\subsection{Atmospheric scale set by \boldmath$\mu_{1,2,3}$ -- Solar scale set by $B_i$}

\begin{figure}
\begin{center} \includegraphics[width=2in]{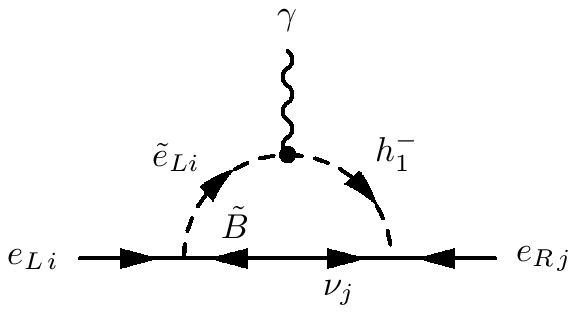} \hspace{10pt} 
\includegraphics[width=2in]{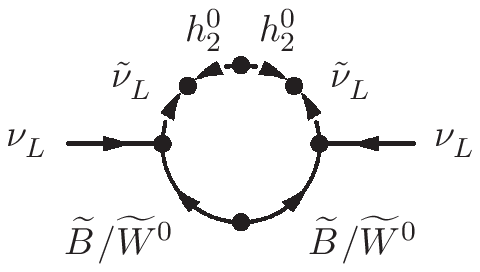}
\caption{\it Feynman diagrams generated when the bilinear terms in the
superpotential and supersymmetry breaking terms, $\mu_{i}$ and $B_{j}$ 
respectively, are non-zero}
\label{dia:b}
\end{center}
\end{figure}

\begin{figure}
{\begin{center}
	\includegraphics[width=3.25in]{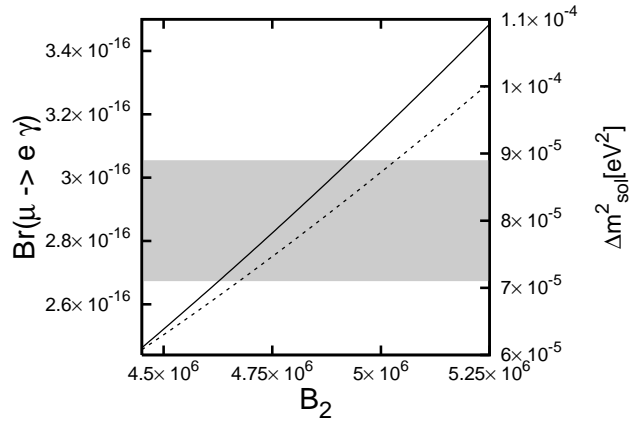}   
\end{center}}
\caption{\it $\tau \rightarrow \mu \gamma$ and $\Delta m^2_{sol}$ against $B_{2}$}
\label{plot:b2mu}
\end{figure}

The diagrams shown in Fig.~\ref{dia:b} are generated when $\mu_i,B_j \neq 0$. 
In the first diagram of Fig.~\ref{dia:b}, the mixing on the internal fermion is set by $\mu_j$
and the mixing on the scalar line is set by $B_i$.  Again, we note that the mass
of the particle inside the loop for the rare decay diagram is of the same
order of magnitude as of the particle in the radiative correction to the
neutrino mass.  
The contribution to the branching ratio is approximately given by,
\begin{equation}
	\Gamma(l_i \rightarrow l_j \gamma) \approx
	\frac{3}{(4\pi)^2} \frac{|\lambda_{0jj}|^2}{G_F^2 m_i^2} 
	\frac{e^2}{c_w^2} \; 
	\left[ 
\left(\frac{1}{m^2_{H^+}} \right)
	 \left(\frac{B_i}{m^2_{H^+}}\right) 
\left(\frac{\mu_j g v_u}{m_{\chi^0}}\right)
	\right]^2
\Gamma (l_i \rightarrow l_j  \nu_i \bar{\nu}_j) \; .
\end{equation}
As such,
we see that these diagrams are not ruled out by current experimental bounds, as shown in
Fig.~\ref{plot:b2mu}, and are not within reach of upcoming studies.

\subsection{Atmospheric scale set by \boldmath$\lambda$ -- Solar scale
set by $\lambda'$}

In the following sections, both the atmospheric and solar mass scales are set
by radiative corrections.  Again, we can find combinations of parameters which
correctly describe the neutrino sector and also give rise to experimentally
attainable branching ratios for $l \rightarrow l' \gamma$.

We first consider the case in which $\lambda_{133}$ and $\lambda_{233}$ are
varied over the following range,
\begin{equation}
	\lambda_{133}= - \sqrt{2} \lambda_{233} = 5 \times 10^{-5} - 8 \times 10^{-5} \;.
\end{equation}
\begin{figure} 
{\begin{center}
	\includegraphics{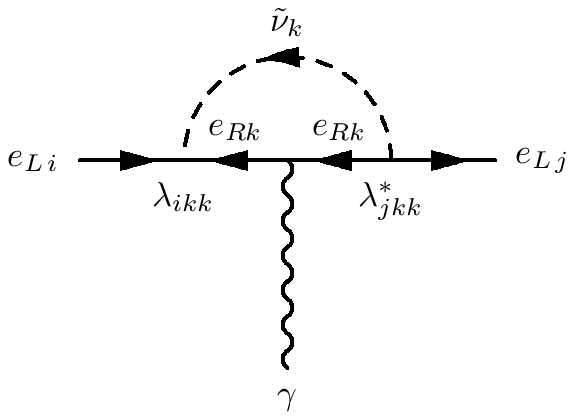} \hspace{30pt} 
\includegraphics{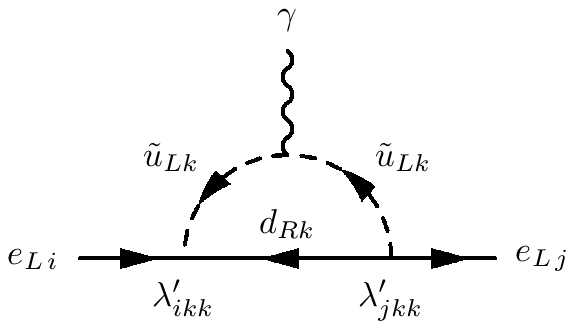} 
	\caption{\it 
Feynman diagrams which contribute to $l_i \rightarrow l_j \gamma$, 
in the case where trilinear lepton number violating couplings are dominant.
	}
\label{dia:lamlam}
\end{center}}
\end{figure}
The ratio ensures the correct mixing between neutrino interaction states is reproduced and the
magnitude sets in atmospheric masses squared difference.  In addition to this,
the contribution of the first diagram in Fig.~\ref{dia:lamlam} to the
branching ratio of $l \rightarrow l'
\gamma$ is approximately given by,

\begin{equation}
	\Gamma(l_i \rightarrow l_j \gamma) \approx
	\frac{3}{(4\pi)^2} \frac{|\lambda_{ikk}|^2 \; |\lambda_{jkk}|^2}{G_F^2} 
	 \;
	\left[ 
	\frac{1}{24} 
	\frac{1}{m^2_{\tilde{\nu}}}
	\right]^2
\Gamma (l_i \rightarrow l_j  \nu_i \bar{\nu}_j) \; .
\end{equation}

The results are given in the upper left panel on
Fig.~\ref{plot:loop1}.  The dashed line (and right hand axis) 
show the atmospheric mass squared difference and the
light grey band shows the values for which $\lambda_{133,233}$ generate an atmospheric mass difference
in agreement with current experimental observations.  The full line (and left
hand axis) show the corresponding branching ratio for $\mu \rightarrow e \gamma$.
The resulting branching ratio is well below current or future experimental sensitivity.

To generate the solar mass squared difference, $\lambda_{1kk,2kk,3kk}$ are
varied in the following hierarchy, ensuring the resulting mixing matrix takes
the form observed by experiment,
\begin{equation}
	\lambda'_{1kk} = \frac{\lambda'_{2kk}}{\sqrt{2}} 
	= - \frac{\lambda'_{3kk}}{\sqrt{3}} \;.
\end{equation}
The second diagram in Fig.~\ref{dia:lamlam}, produces a contribution to the
$l_i \rightarrow l_j \gamma$ branching ratio of approximately,

\begin{equation}
	\Gamma(l_i \rightarrow l_j \gamma) \approx
	\frac{3}{(4\pi)^2} \frac{|\lambda'_{ikk}|^2 \; |\lambda'_{jkk}|^2}{G_F^2} 
	 \;
	\left[ 
	\frac{1}{3} 
	\frac{1}{m^2_{\tilde{u}}}
	\right]^2
\Gamma (l_i \rightarrow l_j  \nu_i \bar{\nu}_j) \; .
\end{equation}

The results for the three possible cases, that is $k=1,2,3$, are shown in the
remaining panels of Fig.~\ref{plot:loop1}.  We note that while $k=2,3$ are
well below current or planned experimental sensitivity, the scenario in which 
$\lambda'_{111,211,311}$ generate the solar mass squared difference would be
discernible by upcoming experimental studies, although we note that
bounds from $\mu-e$ conversion in nuclei already strongly constrain this set
of parameters~\cite{Huitu:1997bi}.  When the solar mass squared
difference is generated by $\lambda'_{133,233,333}$, bottom right panel of
Fig.~\ref{dia:lamlam}, the branching ratio is of the same order as that given
by $\lambda_{133,233}$, setting the solar scale, and contributes with
opposite sign.  Resulting in the negative gradient shown.

\begin{figure}
{\begin{center}
	\includegraphics[width=3.25in]{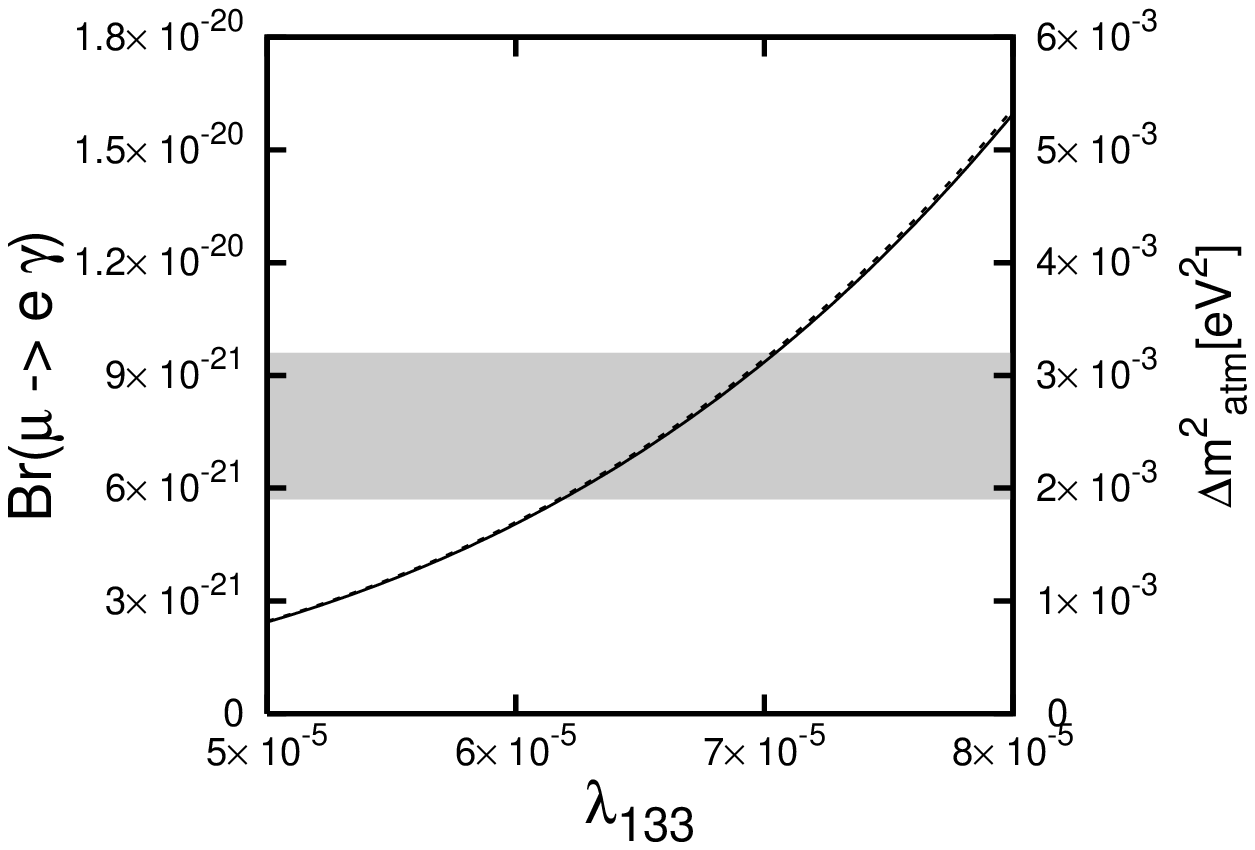} 
	\includegraphics[width=3.25in]{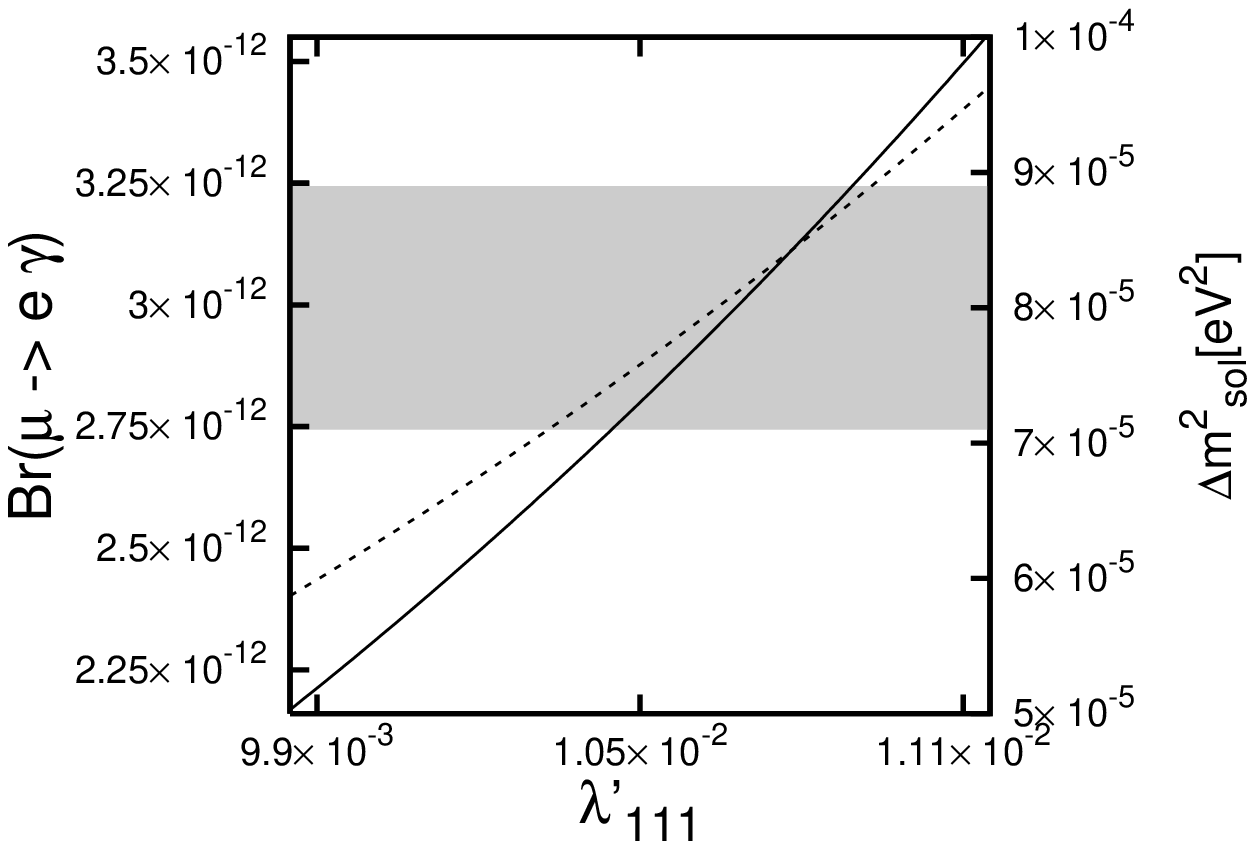}\\
	\includegraphics[width=3.25in]{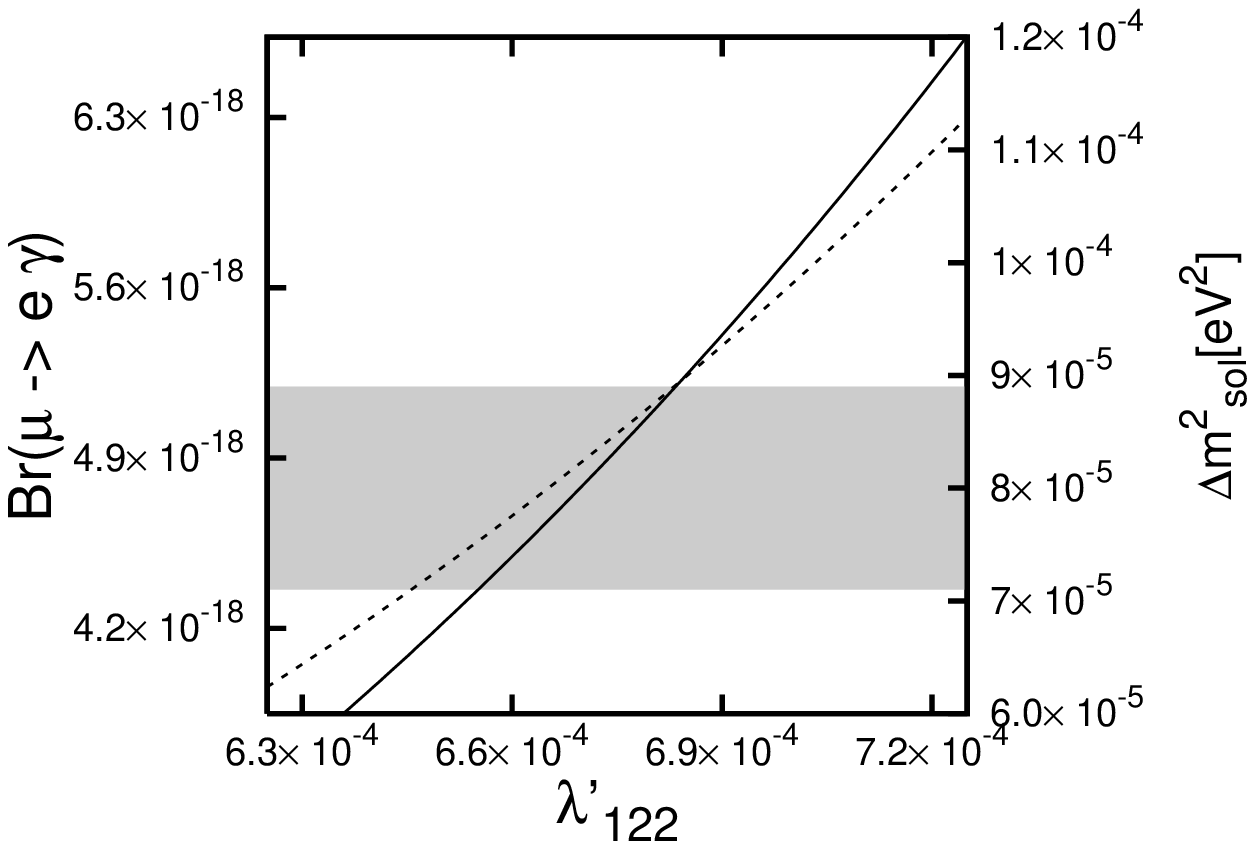}
	\includegraphics[width=3.25in]{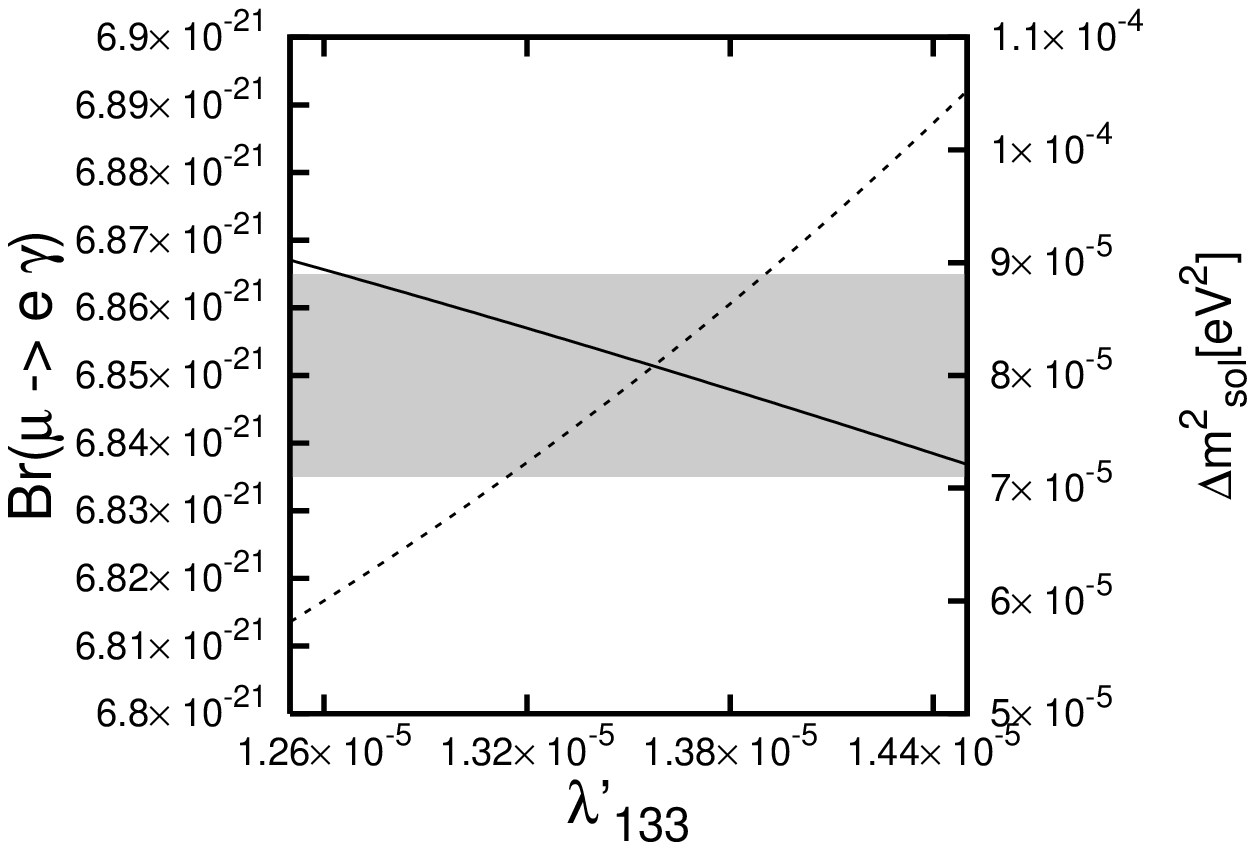}
\end{center}}
\caption{\it
$\mu \rightarrow e \gamma$ and $\Delta m^2_{sol,atm}$ against $\lambda^{(')}$ --
In the top left panel $\lambda_{133,233}$ are varied and the resulting  
branching ratio for $\mu \rightarrow e \gamma$ (full line, left hand axis)
and  $\Delta m^2_{atm}$ (dashed line, right hand axis, grey band showing
current $3\sigma$ band from experiment) are noted.  In the remaining three
panels, $\lambda'_{1kk,2kk,3kk}$ are varied and  Br($\mu \rightarrow e \gamma$)
and $\Delta m^2_{sol}$ are shown.}
\label{plot:loop1}
\end{figure}

\subsection{Atmospheric scale set by \boldmath$\lambda'$ -- Solar scale
set by $\lambda^{(')}$}

Again, both the atmospheric and solar mass scales are set
by radiative corrections.  First, we consider the scenario in which
the atmospheric mass squared difference is set by $\lambda'_{111,211,311}$.
The parameters are given by,
\begin{equation}
		\lambda'_{111} = \frac{\lambda'_{211}}{\sqrt{2}} 
	= \frac{\lambda'_{311}}{\sqrt{3}} \;.
\end{equation}

\begin{figure}
{\begin{center}
	\includegraphics[width=3.25in]{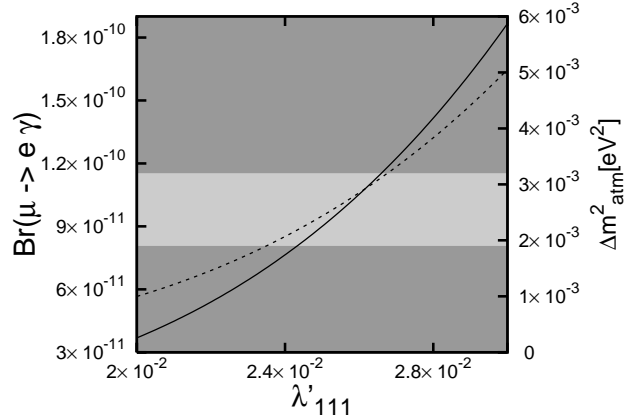} 
\end{center}}
\caption{\it
$\mu \rightarrow e \gamma$ and $\Delta m^2_{atm}$ against $\lambda'_{111}$ --
$\lambda'_{111,211,311}$ are varied and the resulting  
branching ratio for $\mu \rightarrow e \gamma$ (full line, left hand axis, dark grey 
area showing values excluded by experimental searches) 
and  $\Delta m^2_{atm}$ (dashed line, right hand axis, light grey band showing
current $3\sigma$ band from experiment) are noted.}
\label{plot:loopatm}
\end{figure}

The resulting atmospheric mass squared difference and resulting branching
ratio for $\mu \rightarrow e \gamma$ are shown in Fig.~\ref{plot:loopatm},
from which it can be seen that the parameter space which brings about the correct
atmospheric mass difference is already ruled out by the rare decay searches.

In a similar fashion, we consider the scenario in which the atmospheric mass squared
difference is set by $\lambda'_{133,233,333}$, as follows,
\begin{equation}
		\lambda'_{133} = \frac{\lambda'_{233}}{\sqrt{2}} 
	= \frac{\lambda'_{333}}{\sqrt{3}} \;.
\end{equation}
The results are given in the upper left panel of Fig.~\ref{plot:loop2}.  In
this case, we note that the values of $\lambda'_{133,233,333}$ which give the
correct value for the neutrino mass, generate negligible rates for $\mu
\rightarrow e \gamma$.

The solar mass squared difference must now be generated.  It can be set either by $\lambda_{133,233}$ or
by a different set of $\lambda'$ couplings.  First we examine $\lambda'_{1kk,2kk,3kk}$, which are
varied as follows,
\begin{equation}
		\lambda'_{1kk} = \frac{\lambda'_{2kk}}{\sqrt{2}} 
	=- \frac{\lambda'_{3kk}}{\sqrt{3}} \;.
\end{equation}
The upper right and bottom left panels show the results for these parameters.
Again, we note that although the scenario in which the solar scale is set by
$\lambda'_{122,222,322}$ (bottom left panel) is not within experimental sensitivity,
$\lambda'_{111,211,311}$ (top right panel) is close to the current bounds and would be 
seen by searches planned for the near future.  

The final possibility is that the solar mass squared difference be set by $\lambda_{133,233}$.  The parameters
are varied in the following hierarchy,
\begin{equation}
	\lambda_{133}= - \sqrt{2} \lambda_{233} \;.
\end{equation}
The results, given in the bottom right panel of Fig.~\ref{plot:loop2}, show
that this scenario is well below both current and future experimental bounds.

\begin{figure}
{\begin{center}
	\includegraphics[width=3.25in]{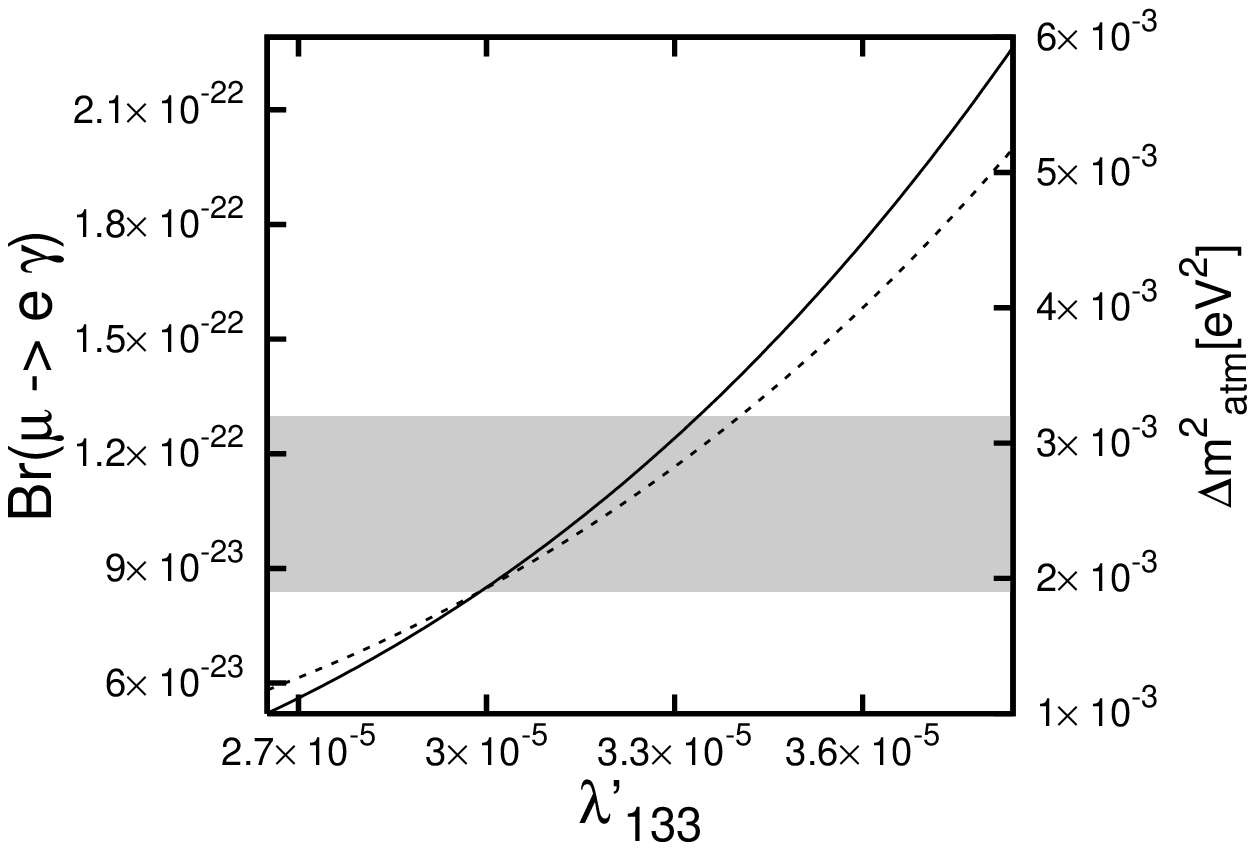} 
	\includegraphics[width=3.25in]{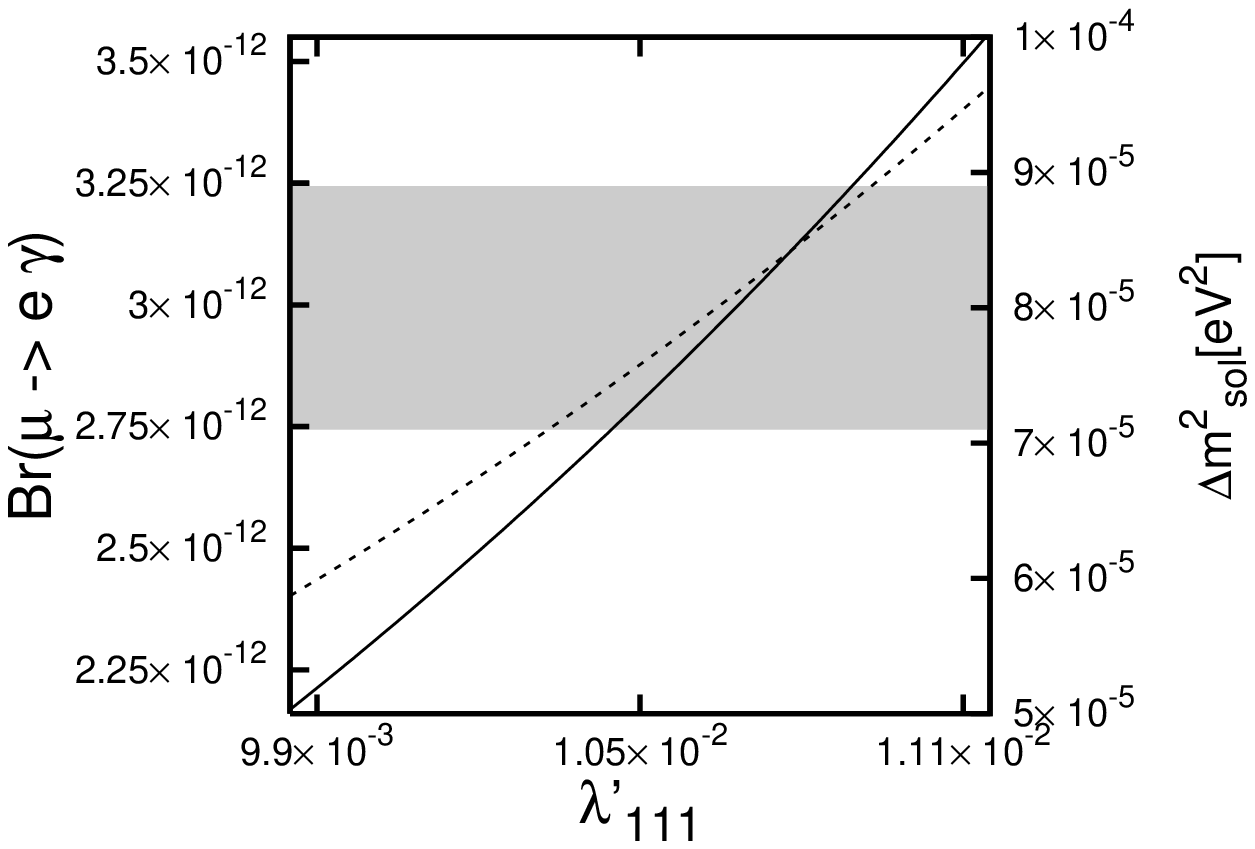}\\
	\includegraphics[width=3.25in]{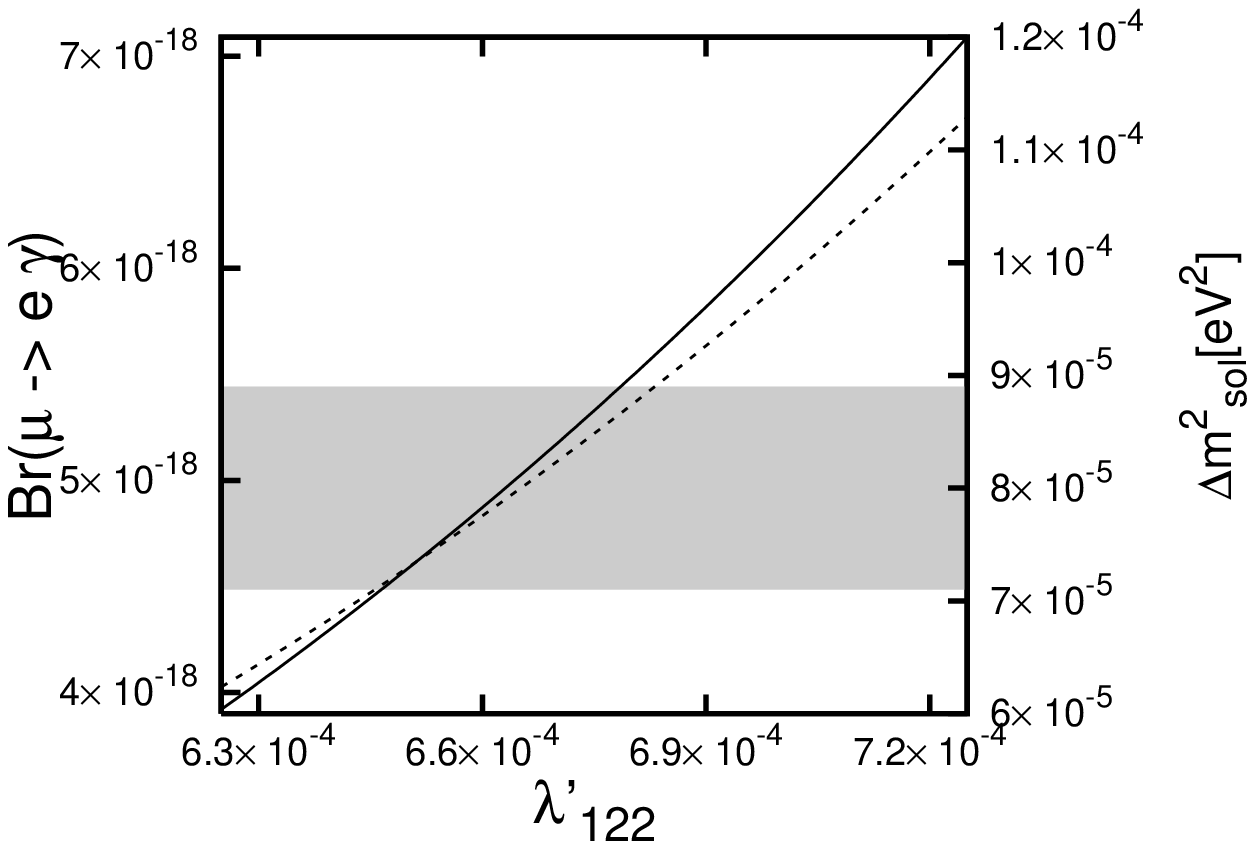}
	\includegraphics[width=3.25in]{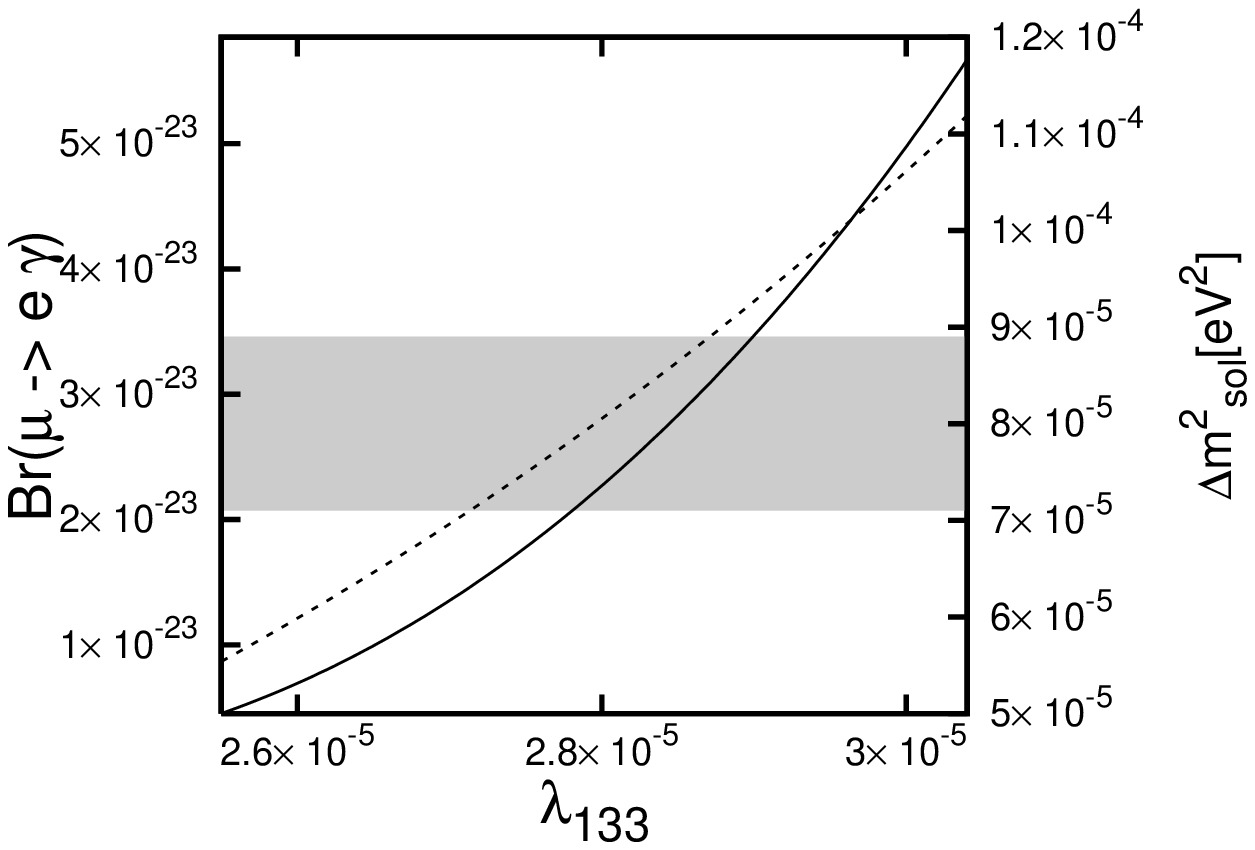}
\end{center}}
\caption{\it
$\mu \rightarrow e \gamma$ and $\Delta m^2_{sol,atm}$ against
$\lambda^{(')}$ --
In the top left panel $\lambda'_{133,233,333}$ are varied and the resulting  
branching ratio for $\mu \rightarrow e \gamma$ (full line, left hand axis)
and  $\Delta m^2_{atm}$ (dashed line, right hand axis, grey band showing
current $3\sigma$ band from experiment) are noted.  In the top right and
bottom left panels, $\lambda'_{111,211,311}$ and $\lambda'_{122,222,322}$ 
are varied, respectively, and the $\mu \rightarrow e \gamma$
and $\Delta m^2_{sol}$ are shown.  In the bottom right hand panel
$\lambda_{133,233}$ are varied and the $\mu \rightarrow e \gamma$
and $\Delta m^2_{sol}$ are given.
}
\label{plot:loop2}
\end{figure}

\section{Benchmarks}

Benchmark scenarios for studying R-parity violating models have been
suggested~\cite{Allanach:2006st}, for which the mass spectrum, nature of the
lightest supersymmetric particle and decays have been studied.  The benchmarks
presented in~\cite{Allanach:2006st} are selected as they produce interesting
signatures at future colliders and are constrained by measurements of
$(g-2)_\mu$, the $b \rightarrow s \gamma$ decay and mass bounds from direct
particle searches.

In addition to these benchmarks, we suggest some other interesting scenarios.
Our motivation being that neutrino data is known and
we demand that the model correctly reproduces these results.  As there are
numerous combinations of lepton number violating parameters which satisfy this
requirement, we consider scenarios for which the upcoming $l \rightarrow l' \gamma$ searches
will constrain the Lagrangian parameters.

The following combinations of lepton number violating parameters are
considered, where all R-parity conserving parameters are set at the SPS1a
benchmark point,

\begin{itemize}
	\item{ {\it Benchmark Scenario 1} -- 
$$	\mu_1 = \frac{\mu_2}{\sqrt{2}} = \frac{\mu_3}{\sqrt{3}} =
1.47 \textup{MeV} \;\;,\;\; \lambda_{211} = 7.4 \times 10^{-4}  $$}
	\item{ {\it Benchmark Scenario 2} -- 
$$	\mu_1 = \frac{\mu_2}{\sqrt{2}} = \frac{\mu_3}{\sqrt{3}} =
1.47 \textup{MeV} \;\;,\;\; \lambda_{311} = 3.7 \times 10^{-2}  $$}
	\item{ {\it Benchmark Scenario 3} -- 
$$ \lambda_{133}= - \sqrt{2} \lambda_{233} = 6.5 \times 10^{-5} 
\;\;,\;\; \lambda'_{111} = \frac{\lambda'_{211}}{\sqrt{2}} 
= - \frac{\lambda'_{311}}{\sqrt{3}} = 1.05 \times 10^{-2}  $$}
	\item{ {\it Benchmark Scenario 4} -- 
$$	\lambda'_{133} = \frac{\lambda'_{233}}{\sqrt{2}} 
= \frac{\lambda'_{333}}{\sqrt{3}} = 3.25 \times 10^{-5} 
 \;\;,\;\;  \lambda'_{111} = \frac{\lambda'_{211}}{\sqrt{2}} 
 = - \frac{\lambda'_{311}}{\sqrt{3}} = 1.05 \times 10^{-2} \; .$$}
\end{itemize}

In the first two benchmark scenarios the observed flavour oscillations of
atmospheric neutrinos are driven by the bilinear lepton number violating
term in the superpotential giving rise to a mass difference at tree level.
For this to occur, the $\mu_i$ parameters are of the order 1 MeV.  In
Benchmark Scenario 1, the solar mass squared difference is then generated by
setting $\lambda_{211} = 7.4 \times 10^{-4}$.  Merely for comparison, we note
that this is approximately $\lambda_{211} \sim 25 \, y_e$, where $y_e$ is the
Yukawa coupling associated with a given particle, in this case being the
electron.  This will give rise to
branching ratios for $\mu \rightarrow e \gamma$ which can be probed by upcoming
experimental studies.  In Benchmark 2 the solar mass squared difference is
determined by $\lambda_{311}=3.7 \times 10^{-2} \sim 1200 \, y_e$.  This
combination of parameters will generate a branching ratio for $\tau
\rightarrow e \gamma$ which may be probed by future studies of $\tau$ decays.

In the Benchmark Scenarios 3 and 4 both the atmospheric and solar mass squared
differences are set by radiative corrections.  The bilinear lepton number
violating terms are set to zero, and the neutrinos are all massless at tree-level.
In Benchmark Scenario 3, $ |\lambda_{133,233}| 
\sim 3 y_e$ set the atmospheric mass difference and 
$|\lambda'_{111,211,311}|  \sim 40 \, y_d$ sets the solar mass
squared difference.  In Benchmark Scenario 4, 
$|\lambda'_{133,233,333}|  \sim 0.1 \, y_d$ sets
the atmospheric scale and 
$|\lambda'_{111,211,311}| \sim 40 \, y_d$
 gives the solar scale.  In both Benchmarks 3 and 4, $\mu \rightarrow e
 \gamma$ would give branching ratios which will be observed by future
 experimental studies.

\section{Conclusions}
\label{sec:conc}

That lepton flavour violating decays of charged particles have not been observed is worthy of
note.  The suppressed branching ratios arise automatically in the Standard
Model, but this is not the case in Supersymmetric extensions where it already
puts strong bounds on certain parameters.  In this paper, we have examined the
effects of lepton number violating couplings on these branching ratios.   Combinations of parameters 
which describe the neutrino sector were chosen to be examined 
and their subsequent effect on the rare decays considered. 

We first investigated the case in which R-parity conserving parameters are set to
the SPS1a benchmark point and the bilinear lepton number violating term in the superpotential,
$\mu_{i}$ generates the atmospheric mass difference.  We showed that the values of $\mu_{i}$
which correctly describe the atmospheric mass difference and mixing angles are not
ruled out by the current bounds on $\tau \rightarrow \mu \gamma$, $e \gamma$ or $\mu \rightarrow e \gamma$.
As such, the bounds from lepton decays are less stringent than the bounds from neutrino data.   
We then considered the case in which a further lepton number violating parameter
correctly reproduces the solar mass difference and considered the combined effect on the $l \rightarrow l' \gamma$ decays.
We note that in this scenario these decays can
impose constraints on one of the trilinear lepton number violating parameters in the
superpotential, $\lambda$.  We considered all the examples in which the $\lambda$ coupling has symmetric final indices, 
which generate the solar neutrino mass with just one non-zero coupling.  $\lambda_{211}$ and $\lambda_{122}$ are excluded by experimental searches for $\mu \rightarrow e \gamma$; $\lambda_{311}$, $\lambda_{133}$, $\lambda_{332}$, $\lambda_{233}$
are not.  We note, however, that the branching ratios are sensitive to the masses of the scalar particles in the loop.
As such, in scenarios where the scalar masses are heavier than those in SPS1a, the branching ratios can be greatly suppressed.

In this scenario, the limits on $l \rightarrow l' \gamma$ do not place useful constraints on
$\lambda'$, or the bilinear lepton number violating terms in the supersymmetry
breaking part of the Lagrangian, $B_i$;
generally, the current constraints from the neutrino sector are stronger.

Second, we considered the scenario in which trilinear lepton number violating couplings are dominant.  
We set all bilinear lepton number violating couplings to zero, and again set R-parity conserving parameters
to the SPS1a benchmark point.  In this scenario, both neutrino mass scales are determined by radiative corrections
and in order to generate the correct mixing matrix in the lepton sector, more than one lepton number violating coupling
must be non-zero for each mass scale.  Because of this, diagrams which contribute to $l \rightarrow l' \gamma$ are generated.
We note that limits already exist when $\lambda'_{111,211,311}$ are used to generate mass differences in this scenario.  In
general however, for $\lambda^{(')}_{x22,x33}$ the constraints from the neutrino masses are stringent.  

\vspace*{0.6cm}

\noindent {\bf Acknowledgements}

\noindent I would like to thank Athanasios Dedes and Janusz Rosiek for their help and
support in producing this work.  I would also like to acknowledge the 
award of a PPARC studentship.

\pagebreak
\appendix

\section{Appendix}
In this section, we present the Feynman Rules for the \LMSSM.  The mixing
matrices, $Z$, diagonalise the mass matrices of the model and determine the
amount of interaction eigenstate in each mass eigenstate.  The full mass
matrices and definitions of the mixing matrices are presented in
Ref.~\cite{Dedes:2006ni}.   
\vspace{-5pt}
\subsection{Neutral Scalar - Charged Fermion - Charged Fermion interactions}

\noindent
\iin{\Km{p}{}}{ \Kp{r}{}}{\Hz{q}}{
- \frac{ie}{s_W}\isqt \ZR{1q}  \Zms{1p}  \Zp{2r} 
-  \frac{ie}{s_W}\isqt \ZR{(2+\alpha)q}  \Zp{1r}  \Zms{(2+\alpha)p}  
}{
+i \lam{\alpha \beta j} \isqt \ZR{(2+\beta)q}   \Zms{(2+\alpha)p}  \Zp{(2+j)r}
}{}{}{}\\
\iin{\Km{p}{}}{\Kp{r}{}}{\Az{q}}{ 
-  \frac{e}{s_W} \isqt \ZA{1q}  \Zms{1p}  \Zp{2r}  
-  \frac{e}{s_W} \isqt \ZA{(2+\alpha)q}  \Zp{1r}  \Zms{(2+\alpha)p} 
}{
- \lam{\alpha \beta j} \isqt \ZA{(2+\beta)q}  \Zms{(2+\alpha)p} \Zp{(2+j)r}
}{}{}{}\\
\vspace{-60pt}
\subsection{
Charged Scalar - Neutral Fermion - Charged Fermion interactions
}

\noindent
\iisi{\Km{p}{}}{\Kz{r}{}}{\Hp{q}}{
-i \frac{e}{s_W}\ZH{(2+\alpha)q}  \Zms{1p}  \ZN{(4+\alpha)r}  
-i \lam{\alpha \beta j} \ZH{(5+j)q}  \Zms{(2+\beta)p}  \ZN{(4+\alpha)r} 
}{
+i  \frac{e}{\sqrt{2} c_W}  \ZH{(2+\alpha)q} \Zms{(2+\alpha)p}  \ZN{1r} 
+i \frac{e}{\sqrt{2} s_W}\ZH{(2+\alpha)q} \ZN{2r}  \Zms{(2+\alpha)p} }{}{}{}\\
\iiso{\Kp{p}{}}{\Kz{r}{}}{\Hps{q}}{
-i \lam{\alpha \beta j} \ZHs{(2+\beta)q}  \Zp{(2+j)p}  \ZN{(4+\alpha)r} 
-i  \sqrt{2} \frac{e}{c_W}  \ZHs{(5+i)q}  \Zp{(2+i)p} \ZN{1r} 
}{
-i  \frac{e}{\sqrt{2} c_W}  \ZHs{1q}  \Zp{2p}  \ZN{1r} 
-i \frac{e}{s_W}\ZHs{1q}  \Zp{1p} \ZN{3r}  
}{
-i \frac{e}{\sqrt{2} s_W} \ZHs{1q} \ZN{2r} \Zp{2p}}{}{}{}\\

\subsection{
Squark - Charged Fermion - Quark interactions
}

\noindent
\iisi{\Km{p}{}}{\dR{k}{y}{}}{\su{q}{y}}{
+i \lamp{\alpha ij} \Zsu{iq}  \Zms{(2+\alpha)p}  \ZdR{jk} 
}{}{}{}{}\\
\iiso{\Kp{p}{}}{\dL{j}{y}{}}{\sus{q}{y}}{
-i \frac{e}{s_W} \Zsus{iq} \Zp{1q} \ZdLs{ij} 
+i \YU{ik} \Zsus{(3+k)q} \ZdLs{ij}  \Zp{2q}
}{}{}{}{}\\
\iisi{\uR{k}{y}{}}{\Kp{p}{}}{\sd{q}{y}}{
+i \YU{ij} \Zsds{iq}  \ZuRs{jk}   \Zp{2p} 
}{}{}{}{}\\
\iiso{\Km{p}{}}{\uL{k}{y}{}}{\sds{q}{y}}{
+i \lamp{\alpha ij} \Zsd{(3+j)q}  \Zms{(2+\alpha)p}  \ZuL{ik} 
-i \frac{e}{s_W} \Zsd{iq} \Zms{1p}  \ZuL{ik}
}{}{}{}{}\\

\pagebreak

\end{document}